\documentclass[journal]{new-aiaa}  
\usepackage[utf8]{inputenc}
\usepackage{graphicx}
\usepackage{tabularx}
\usepackage{caption}
\usepackage{subcaption}
\usepackage{amsmath}
\usepackage[version=4]{mhchem}
\usepackage{siunitx}
\usepackage{longtable,tabularx}
\usepackage{multirow}   
\usepackage{color}      
\usepackage{soul}       
\usepackage{enumitem}
\newcommand\nnfootnote[1]{%
  \begin{NoHyper}
  \renewcommand\thefootnote{}\footnote{#1}%
  \addtocounter{footnote}{-1}%
  \end{NoHyper}
}

\title{Large-Eddy Simulation and Performance Analysis of a Continuous-Turbine-Burner Stage}

\author{Yalu Zhu\footnote{Assistant Specialist, Department of Mechanical and Aerospace Engineering, yalu.zhu@uci.edu (Corresponding Author), Member AIAA.}, 
Feng Liu\footnote{Professor, Department of Mechanical and Aerospace Engineering, Fellow AIAA.} 
and William A. Sirignano\footnote{Distinguished Professor, Department of Mechanical and Aerospace Engineering, Honorary Fellow AIAA.}} 
\affil{University of California, Irvine, Irvine, CA, 92697-3975}

\begin{document}

\nnfootnote{Part of this work was presented as Paper 2025-1129 at the AIAA SciTech 2025 Forum, Orlando, FL, January 6–10, 2025.}

\maketitle
\singlespacing

\begin{abstract}
This paper reports the first large-eddy simulation of chemically reacting flow in a turbine stage to analyze the influence of fuel injection and combustion on its aerodynamic and thermodynamic performance. 
Two reacting cases---with four and sixteen fuel injectors at the inlet for each stator passage---are computed and compared against two nonreacting cases, one with four fuel injectors and the other without.
The analyses indicate viability for the continuous-turbine-burner (CTB) concept, which offers the potential of significant increase of specific power/thrust of a gas-turbine engine without loss of efficiency. Fuel injection and combustion have minimal influence on the total-pressure loss. 
Compared with the baseline nonreacting case, the stage work per unit mass increases by 8.5\% and 11.5\% in the two reacting cases, while the residual work rises by 17.3\% and 16.0\%, respectively. 
The two reacting cases exhibit a 14.5\% increase of overall work and a thermal efficiency of 44\% for the fuel injection.  
Local high temperature on the rotor blade is suppressed by using a more uniform spanwise distribution of fuel injectors.
The work extraction process of a CTB is analyzed from both thermodynamic and mechanical views.
\end{abstract}

\section*{Nomenclature}
{\renewcommand\arraystretch{1.0}
\noindent\begin{longtable*}{@{}l @{\quad=\quad} l@{}}
\addtocounter{table}{-1}
$C_p$ & specific heat capacity at constant pressure \\
$E$ & absolute total energy, $E= H - p/\rho$ \\
$H$ & absolute total enthalpy, $H = h +1/2 \mathbf{V} \cdot \mathbf{V}$ \\
$h$ & sensible enthalpy (excluding enthalpy of formation) \\
$h^0$ & enthalpy of formation at reference temperature $T_{\mathrm{ref}}$ \\
$\mathbf{I}$ & Kronecker tensor \\
$\mathbf{j}$ & diffusive flux \\
$\dot{m}$ & mass flow rate \\
$N$ & total number of species \\
$\mathrm{Pr}$ & Prandtl number \\
$P$ & power of turbine \\
$p$ & pressure \\
$\dot{Q}$ & heat release rate \\
$\mathbf{q}$ & heat flux \\
$R$ & gas constant \\
$\mathbf{r}$ & radius vector from turbine rotation axis \\
$r$ & increase rate of work due to combustion \\
$\mathrm{Sc}$ & Schmidt number \\
$T$ & temperature \\
$T_{\mathrm{ref}}$ & reference temperature, $T_{\mathrm{ref}} = 298.15 \, \mathrm{K}$ \\
$t$ & time \\
$\mathbf{V}$ & absolute velocity vector \\
$W$ & specific work of turbine \\
$x$ & axial coordinate from stator leading edge \\
$Y$ & mass fraction \\

$\beta$ & flow angle \\
$\eta_f$ & thermal efficiency of fuel burning \\
$\mu$ & viscosity coefficient \\
$\rho$ & density \\
$\boldsymbol{\tau}$ & viscous stress tensor \\
$\mathbf{\Omega}$ & angular velocity of rotor \\
$\dot{\omega}$ & mass production rate of species \\

$(\cdot)_i$ & quantity of species $i$ \\
$(\cdot)_p$ & potential quantity at turbine-stage outlet \\
$(\cdot)_T$ & quantity of turbulence \\
$(\cdot)_t$ & quantity of turbine stage \\
$(\cdot)_0$ & quantity in stagnation state \\
\end{longtable*}}

\section{Introduction}
The gas turbine plays an essential role in power generation, aircraft propulsion, and ship propulsion. In a conventional gas-turbine engine, designers usually seek to prevent hot streaks in the upstream combustor from propagating into the downstream turbine passage, thereby reducing the challenge of heat transfer on the turbine blade. 
However, the thermal-cycle performance analyses of the turbine-burner type of gas-turbine engines proposed by Sirignano and Liu \cite{sirignano1999performance, liu2001turbojet} showed that intentional augmented burning within the turbine passage can shorten and lighten the turbine, reduce specific fuel consumption, and increase specific thrust.  
Three configurations were studied: (1) continuous turbine-burner (CTB) \cite{andriani1999jet, andriani2002off, chiu2005performance, sirignano1999performance, liu2001turbojet}, where combustion continues throughout the turbine stage as the flow is accelerating, aiming to approximate an isothermal expansion inside the whole turbine; (2) interstage turbine-burner (ITB) \cite{liu2001turbojet, liew2005parametric, liew2006performance, yin2020review, jia2023design, li2025off}, in which combustors are placed in the transition duct between the high-pressure turbine (HPT) and the low-pressure turbine (LPT); and (3) an intermediate configuration, called the multiple interstage burners (MIBs) \cite{liu2001turbojet, chen2004gas}, in which the flow is reheated within multiple axial gaps between  the stators and rotors of the turbine.
Spytek Aerospace developed prototype engines based on the ITB concept and achieved significant power increases and high efficiency in engine tests \cite{spytek2019small, Spytekaeropace}.

The ITB engines tested by Spytek \cite{spytek2019small, Spytekaeropace} are based on isolated combustors inserted between two turbine stages. Liu and Sirignano\cite{liu2001turbojet} proposed a  more compact ITB design in which the turbine-burner is integrated with the turbine nozzle (stator) blade row for reduced length and weight of the engine.
Furthermore, they showed that the CTB design, where combustion is allowed to continue through the rotor blade row, would provide even higher specific power/thrust and efficiency compared to the ITB and the MIB designs.
However, both the integrated ITB and the CTB designs face challenges associated with the reacting flow inside the turbine, more so with the CTB design where combustion continues through the rotating blade row.

The flow in a turbine passage is subjected to strong pressure gradients generated by the blade profiles, causing it accelerate from subsonic to supersonic within a very short distance. In addition, large gradients of temperature, velocity, and species concentration occur on the fuel-oxidizer interface due to mixing and combustion. Moreover, strong interactions between turbulence and chemical reactions happen within the flames.
These features lead to many challenges in sustaining reacting flow inside a turbine-burner \cite{sirignano2012turbine}, including: ignition and flameholding in a highly accelerating flow; achieving complete fuel-oxidizer mixing and combustion within a short residence time; maintaining acceptable aerodynamic loading on the rotor blade; mitigating hydrodynamic instability in the mixing layer; and avoiding over-heating of the blades. 

The aforementioned challenges in a turbine-burner cannot be adequately addressed through overall thermal-cycle analyses. Instead, numerical simulations of reacting flows under strong favorable pressure gradients or within turbine-like configurations are required, given the limited feasibility of conducting experimental measurements prior to the design of a turbine–burner system.
Sirignano and Kim \cite{sirignano1997diffusion} obtained the similarity solution for a diffusion flame in the two-dimensional (2D), laminar, steady, compressible mixing layer with constant pressure gradient. 
Fang et al. \cite{fang2001ignition} extended the study to include cases with arbitrary pressure gradients by solving the boundary-layer equations. 
Mehring et al. \cite{mehring2001ignition} further extended the laminar boundary-layer computation to a turbulent one. 
To capture the combustion process in line with multi-step finite-rate kinetics, Walsh et al. \cite{walsh2025turbulent, walsh2026performance} used a flamelet progress variable model to simulate the accelerating turbulent mixing layer using the boundary-layer approximation and by solving the 2D Reynolds-averaged Navier-Stokes (RANS) equations.
Cai et al. \cite{cai2001ignition, cai2001combustion} investigated the turbulent reacting flow in a curved duct mimicking a turbine passage by solving the 2D RANS equations. 
Cheng et al. \cite{cheng2007nonpremixed, cheng2008nonpremixed, cheng2009reacting} computed the unsteady transitional evolution of reacting mixing layers in curved ducts by solving the 2D Navier-Stokes (N-S) equations. 
In these simulations, a mixing layer under pressure gradient or a curved convergent-divergent duct is used to model the accelerating diffusion flame in a turbine-burner. 

To evaluate the effects of combustion on the aerodynamic and thermodynamic performance of a turbine-burner, realistic turbine configurations must be considered.
Zhu et al. \cite{zhu2024numerical} applied the RANS equations with a finite-rate chemistry model to simulate the turbulent reacting flow in a highly-loaded turbine cascade. 
Walsh et al. \cite{walsh2026flamelet} analyzed the same cascade using a flamelet-based chemistry model. 
Zhu et al. \cite{zhu2025_turbine_burner} simulated the turbulent reacting flow in an experimental low-speed turbine stage using large-eddy simulation (LES).
To verify the viability for the turbine-burner concept, a practical three-dimensional (3D) turbine configuration is required, rather than simplified 2D cascades or idealized experimental turbines. This is the first motivation of the present study.

The existing computations of the reacting flows with large pressure gradients and in turbine configurations are typically based on the boundary-layer approximation \cite{fang2001ignition, mehring2001ignition, walsh2025turbulent} or 2D N-S equations \cite{cai2001ignition, cai2001combustion, cheng2007nonpremixed, cheng2008nonpremixed, cheng2009reacting, zhu2024numerical, walsh2026flamelet, walsh2026performance}. Influence of turbulence is not incorporated in these simulations \cite{sirignano1997diffusion, fang2001ignition, cheng2007nonpremixed, cheng2008nonpremixed, cheng2009reacting} or simply modeled by a Reynolds-averaged approach \cite{mehring2001ignition, walsh2025turbulent, cai2001ignition, cai2001combustion, zhu2024numerical, walsh2026flamelet, walsh2026performance}. 
In order to evaluate the influence of chemical reactions on turbine performance, high-fidelity numerical simulations that solve the full 3D N-S equations with appropriate chemistry and turbulence models must be applied. LES is one such approach, as it resolves the most large-scale structures and turbulence-reaction interactions in turbulent reacting flows \cite{zhu2025_turbine_burner}. This is the second motivation of the present study.

A high-fidelity simulation of a turbine-burner provides detailed flow information that can be used not only to assess the aerodynamic performance of the turbine-burner itself, but also to evaluate its influence when the resulting performance parameters are integrated into the engine's thermodynamic cycle or, at least, into the thermodynamic process within the entire turbine section. This performance analysis is expected to be more reliable than overall thermal-cycle analyses \cite{sirignano1999performance, liu2001turbojet}, which rely on simplified lumped thermodynamic and combustion models for turbine-burner systems.
Such analysis can provide guidance to strategies for improving the aerodynamic and thermodynamic performance of a turbine-burner. This is the third motivation of the present study. 

To address the three aforementioned motivations related to turbine–burners, the present study performs 3D LES simulations of the turbulent reacting flow in a practical turbine stage and then analyzes the influence of fuel injection and combustion on the aerodynamic and thermodynamic performance of the stage. 
The 3D compressible N-S equations, along with a one-step chemistry model and a LES turbulence model, are solved in multi-block grids by using a finite-volume method. 
The remainder of the paper is organized as follows. 
The governing equations and numerical methods are presented in Sec. \ref{sec:numerical_methods}. 
The unsteady reacting flow fields in the stage are discussed in Sec. \ref{sec:reacting_flow}.
The effects of combustion on the aerodynamic and thermodynamic performance of the turbine are evaluated in Secs. \ref{sec:aerodynamic_performance} and \ref{sec:thermodynamic_performance}, respectively.
Concluding remarks are given in Sec. \ref{sec:conclusion}.

\section{Numerical Methods} \label{sec:numerical_methods}

\subsection{Governing Equations}
The governing equations are the 3D, unsteady, compressible, multi-component N-S equations. A turbine stage consists of a stator followed by a rotor. By expressing the governing equations in the rotating frame of reference while using absolute flow quantities, they are valid for both rotor and stator by setting the angular velocity to the rotor value or zero for the stator \cite{liu2025computational}.
The transport equations for partial density of each species in a mixture with $N$ species, and momentum and energy of the mixture are
\begin{subequations} \label{eq:navier-stokes}
    \begin{align}
        \frac{\partial{(\rho Y_i)}}{\partial t} + \nabla \cdot (\rho Y_i \mathbf{V^\prime}) & = -\nabla \cdot \mathbf{j}_i + \dot{\omega}_i, \ i = 1, \ 2, \ ..., \ N   \label{eq:species} \\
        \frac{\partial(\rho\mathbf{V})}{\partial t} + \nabla \cdot (\rho\mathbf{V}\mathbf{V^\prime}) & = -\nabla p + \nabla \cdot \boldsymbol{\tau} - \rho\mathbf{\Omega} \times \mathbf{V} \label{eq:momentum} \\
        \frac{\partial(\rho E)}{\partial t} + \nabla \cdot (\rho E \mathbf{V^\prime}) & = -\nabla \cdot (p \mathbf{V}) + \nabla \cdot (\mathbf{V} \cdot \boldsymbol{\tau}) - \nabla \cdot \mathbf{q} + \dot Q \label{eq:energy}         
    \end{align}
\end{subequations}
where $\mathbf{V^\prime} = \mathbf{V} - \mathbf{\Omega} \times \mathbf{r}$.
The source term in the momentum equation (\ref{eq:momentum}) appears because of a Coriolis force when observing the time change rate of the absolute velocity in the rotating frame of reference. It performs no work on the rotor.

The perfect-gas law is assumed.
\begin{equation}
    p = \rho R T
\end{equation}
where $R = \sum_{i=1}^{N}{Y_i R_i}$ is the gas constant of the mixture.  

The total energy $E$ is expressed by
\begin{equation}
    E = h - \frac{p}{\rho} + \frac{1}{2} \mathbf{V} \cdot \mathbf{V}
\end{equation}
with
\begin{equation}
    h = \sum_{i=1}^{N}{Y_i h_i}, \ \ h_i = \int_{T_{\mathrm{ref}}}^{T}{C_{p,i}\mathrm{d}T}
\end{equation}
where $C_{p,i}$ is a function of temperature given by the empirical polynomial formula of NASA \cite{mcbride1993coefficients} for each species. 
In this study, $h$ is defined as the sensible enthalpy, excluding the enthalpy of formation $h^0$, whose contribution to the energy equation (\ref{eq:energy}) is taken into account by the heat release rate
\begin{equation} \label{eq:hrr}
    \dot Q = -\sum_{i=1}^{N}{\dot\omega_i h^0_{i}}
\end{equation}

Methane is chosen as the fuel for the turbine-burner for simplicity. The combustion of methane in air is modeled by the Westbrook and Dryer's one-step reaction mechanism \cite{westbrook1984chemical}
\begin{equation} \label{eq:methane_reaction}
    {\rm CH}_4 + 2 {\rm O}_2 + 7.52 {\rm N}_2 \rightarrow {\rm CO}_2 + 2 {\rm H}_2{\rm O} + 7.52 {\rm N}_2 
\end{equation}
where only five species ($N = 5$), i.e., methane ($\rm{CH_4}$), oxygen ($\rm{O_2}$), nitrogen ($\rm{N_2}$), carbon dioxide ($\rm{CO_2}$), and water vapor ($\rm{H_2O}$), are tracked. 
The expression of production rate $\dot\omega_i$ of each species due to chemical reaction is directed to Ref. \cite{zhu2024numerical}. 

The present study is limited to the use of the above simple one-step chemistry model, which does not account for detailed multi-step chemistry and the effect of turbulence on combustion. It does, however, capture the dynamic evolution of the chemical reaction in response to the turbulent fluctuations by resolving the reaction time explicitly \cite{xiong2022_30injectors, xiong2022_82injectors}, unlike models that rely on equilibrium assumptions. With one-step reaction, we are not yet able to track minor species such as NO\textit{x}, but the predictions reported here are expected to be essentially accurate in terms of the thermal and fluid interactions. Improved flamelet models, which consider detailed chemical kinetics, non-equilibrium effects \cite{walsh2025turbulent, walsh2026performance}, proper treatment of turbulence scaling \cite{sirignano2025flamelet}, and influence of both strain rates and vorticity \cite{sirignano2022three, sirignano2022inward, sirignano2024unsteady, hellwig2025three, hellwig2025vortex}, are being developed in parallel to this work. Application of one of those newly developed flamelet models was used to study the reacting flow in a turbine stator in Ref. \cite{walsh2026flamelet}. As these new chemical models become validated and computationally efficient, they will be incorporated in future studies of the turbine-burner.

The transport properties in Eq. (\ref{eq:navier-stokes}) are given by 
\begin{subequations} \label{eq:constitution_relation}
    \begin{align}
        \boldsymbol{\tau} &= 2(\mu + \mu_T) \left[ \mathbf{S} - \frac{1}{3}(\nabla \cdot \mathbf{V} ) \mathbf{I} \right], \ \mathbf{S} = \frac{1}{2} \left[\nabla \mathbf{V} + (\nabla \mathbf{V})^\top \right] \\
        \mathbf{j}_i &= -\left( \frac{\mu}{\mathrm{Sc}_i} + \frac{\mu_T}{\mathrm{Sc}_T} \right) \nabla Y_i \\
        \mathbf{q} &= -\left( \frac{\mu}{\mathrm{Pr}} + \frac{\mu_T}{\mathrm{Pr}_T} \right) \left( \nabla h - \sum_{i=1}^{N}{h_i \nabla Y_i} \right) + \sum_{i=1}^{N}{h_i\mathbf{j}_i} \label{eq:heat_flux}
    \end{align}
\end{subequations}
The Soret and Dufour effects are neglected for LES. The last term in Eq. (\ref{eq:heat_flux}) stands for the energy transport due to mass diffusion of each species with different enthalpy. The wall-adapting local eddy-viscosity (WALE) model \cite{nicoud1999subgrid} is adopted to determine the turbulent viscosity $\mu_T$.
The Schmidt number of species $i$, $\mathrm{Sc}_i$, and the Prandtl number $\mathrm{Pr}$ are set as 1.0 while the turbulent Prandtl number $\mathrm{Pr}_T$ and the turbulent Schmidt number $\mathrm{Sc}_T$ are set as 0.6. 
The molecular viscosity $\mu$ is computed by the mass-weighted summation of molecular viscosity of each species given by Sutherland's law \cite{white2006viscous}
\begin{equation} \label{eq:sutherland_law}
    \mu = \mu_{\mathrm{ref}} \left(\frac{T}{T_{\mathrm{ref}}}\right)^{3/2} \frac{T_{\mathrm{ref}} + S}{T + S}
\end{equation}
The reference temperature $T_{\mathrm{ref}}$, reference viscosity $\mu_{\mathrm{ref}}$, and Sutherland constant $S$ in Eq. (\ref{eq:sutherland_law}) are provided in Table \ref{table:sutherland_constant} for each species.

\begin{table}[htb!]
    \caption{Reference values and constants in Sutherland's law for viscosity}
    \centering
    \label{table:sutherland_constant}
    \begin{tabular}{cccc}
      \hline
      \textit{ }   &$T_{\mathrm{ref}}, \ \mathrm{K}$   &$\mu_{\mathrm{ref}}, \ \mathrm{kg/m/s}$     &  $S, \ \mathrm{K}$ \\
      \hline
      $\rm{CH_4}$  &  273  &  $1.200 \times 10^{-5}$  &  198    \\
      $\rm{O_2}$   &  273  &  $1.919 \times 10^{-5}$  &  139       \\
      $\rm{N_2}$   &  273  &  $1.663 \times 10^{-5}$  &  107       \\
      $\rm{CO_2}$  &  273  &  $1.370 \times 10^{-5}$  &  222       \\
      $\rm{H_2O}$  &  350  &  $1.120 \times 10^{-5}$  &  1064      \\
      \hline
    \end{tabular}
\end{table}

\subsection{Numerical Solver}
An in-house code for simulating steady and unsteady compressible flows for single species within turbomachinery blade rows has been developed, validated, and applied \cite{ zhu2017numerical, zhu2018flow, zhu2018influence, liu2025computational}.
Recently, it has been extended to include solving transport equations for multiple species with varying specific heat capacities and appropriate chemistry models, and verified and validated by 2D steady transonic reacting flows in a mixing layer and a turbine cascade \cite{zhu2024numerical} and 3D unsteady reacting flow in a rocket engine \cite{zhu2026simulation}.
The code solves the multi-component N-S equations together with turbulence models by using a second-order cell-centered finite-volume method based on multi-block structured grids. 
Parallel techniques based on message passing interface (MPI) are adopted to accelerate the computation by distributing grid blocks among CPU processors.

In the present study, the diffusive fluxes are computed using a second-order central scheme, where the flow variable gradients on each cell face in Eq. (\ref{eq:constitution_relation}) are evaluated using the Gauss integration method \cite{liu2025computational} within an auxiliary control volume. 
The convective fluxes are discretized using the Jameson-Schmidt-Turkel (JST) scheme \cite{jameson1981numerical}, a central scheme with second- and fourth-order artificial viscosity terms. The convective flux crossing the cell face $j+1/2$ is evaluated as
\begin{equation} \label{eq:pure-central2}
     \mathbf{F}_{c, j+1/2} = \mathbf{F}_c \left( \frac{\mathbf{W}_{j}+\mathbf{W}_{j+1}}{2} \right) - \mathbf{D}_{j+1/2}
\end{equation}
where $\mathbf{W} = [\rho Y_i, \ \rho \mathbf{V}, \ \rho E]^\top$ is the state vector, $\mathbf{F}_c$ is the convective flux vector, and 
\begin{equation} \label{eq:artificial_dissipation}
     \mathbf{D}_{j+1/2} = \Lambda^J_{j+1/2}\left[ \varepsilon^{(2)}_{j+1/2} \left(\mathbf{W}_{j+1}-\mathbf{W}_j \right) -\varepsilon^{(4)}_{j+1/2} \left(\mathbf{W}_{j+2} - 3\mathbf{W}_{j+1} + 3\mathbf{W}_j - \mathbf{W}_{j-1} \right) \right]
\end{equation}
$\Lambda^J$ is scaled to the spectral radius of the Jacobian of the convective flux vector, providing the appropriate scaling for upwinding. 
The coefficients $\varepsilon^{(2)}_{j+1/2}$ and $\varepsilon^{(4)}_{j+1/2}$ are defined as
\begin{subequations} 
    \begin{align}
        \varepsilon^{(2)}_{j+1/2} &= k^{(2)} \min \left[0.25, \max\left(\gamma_j, \gamma_{j+1} \right) \right] \\
        \varepsilon^{(4)}_{j+1/2} &= \max \left[0, \left(k^{(4)} - \varepsilon^{(2)}_{j+1/2} \right) \right] 
    \end{align}
\end{subequations}
The indicator of discontinuity $\gamma_j$ is given by the maximum of the pressure and partial-density gradients.
\begin{equation}
    \gamma_j = \max \left(\gamma_j^p, \gamma_j^{\rho Y_i} \right) \ \mathrm{with} \ \gamma_j^\phi = \frac{\left| \phi_{j+1} - 2\phi_{j} + \phi_{j-1} \right|}{\phi_{j+1} + 2\phi_{j} + \phi_{j-1}}
\end{equation}
The second-order dissipation coefficient $k^{(2)}$ is set to 0.1 to capture the possible partial-density discontinuities between the fuel jets and air, whereas the fourth-order dissipation coefficient $k^{(4)}$ is assigned a small value, 1/32, to minimize numerical dissipation effects in smooth flow regions.

In the species equation (\ref{eq:species}) and energy equation (\ref{eq:energy}), the source terms exhibit fundamentally different physical properties from the terms of convection and diffusion due to significantly smaller time scales for chemical reactions than for the flow, resulting in strong stiffness in solving the governing equations. 
An operator-splitting method is a natural choice to achieve efficient integration in time for both steady and unsteady problems \cite{mott2000quasi}. In our previous work \cite{zhu2024numerical}, a steady-state preserving splitting scheme for solving steady reacting flows was proposed. It was later extended into unsteady problems \cite{zhu2025_turbine_burner} and subsequently applied to simulate unsteady reacting flows \cite{zhu2026simulation}. The present study uses the extended splitting scheme as the time-integration method.

\subsection{Computational Grid and Cases} \label{sec:configuration_stage}
The NASA/GE Energy Efficient Engine (E\textsuperscript{3}) program \cite{timko1984energy} aimed to develop technologies suitable for energy-efficient turbofans for high-subsonic commercial aircraft of the late 1980s or early 1990s. The first stage of the two-stage, low-throughflow HPT of the E\textsuperscript{3} engine is selected as a representative real turbine stage to evaluate the CTB concept by resolving the unsteady reacting flow within it. The turbine stage consists of a nozzle vane (stator) followed by a rotor. Slightly twisted low-aspect-ratio blades are used for both the stator and the rotor. 
The primary design parameters of the turbine stage are listed in Table \ref{table:turbine_info} \cite{claus2015geometry}. The designed total-to-total pressure ratio of the stage is 2.25. 
The Reynolds numbers, based on the blade chord and exit flow conditions, are estimated to be $1.9 \times 10^6$ for the stator and $1.0 \times 10^6$ for the rotor, respectively.

\begin{table}[htb!]
    \caption{Design parameters of the turbine stage}
    \centering
    \label{table:turbine_info}
    \begin{tabular}{l c c}
      \hline
      \textit{ }            &Stator       &Rotor    \\
      \hline
      Blade number          &46           &76       \\
      Hub radius, mm        &325.76       &323.37   \\
      Casing radius, mm     &365.76       &366.01   \\
      Axial chord length, mm &33.78       &28.7     \\
      Relative inlet angle, $^\circ$  &0  &43.2     \\
      Relative outlet angle, $^\circ$ &74.2 &66.9   \\
      Relative inlet Mach number  &0.109  &0.324    \\
      Relative outlet Mach number &0.878  &0.819    \\
      Tip clearance, mm           &/      &0.41     \\
      Rotation speed, rpm         &/      &12630    \\
      \hline
    \end{tabular}
\end{table}

The blade numbers of the stator and rotor are changed from 46:76 to 38:76 to allow computations on a configuration consisting of one stator passage and two rotor passages with periodic boundary condition in the blade-to-blade (B2B) direction. To preserve the stator solidity, the stator blade is scaled accordingly by a factor of the blade-number ratio: 46/38, following Rai's blade-scaling rule \cite{rai1989three}, resulting an axial chord length of $40.89 \, \mathrm{mm}$. Computational results, although not shown here, indicate that this scaling has negligible effect on the radial (also called spanwise) variation of aerodynamic performance at the stator and rotor outlets. 

\begin{figure}[htb!]
    \centering
    \includegraphics[width=0.45\linewidth]{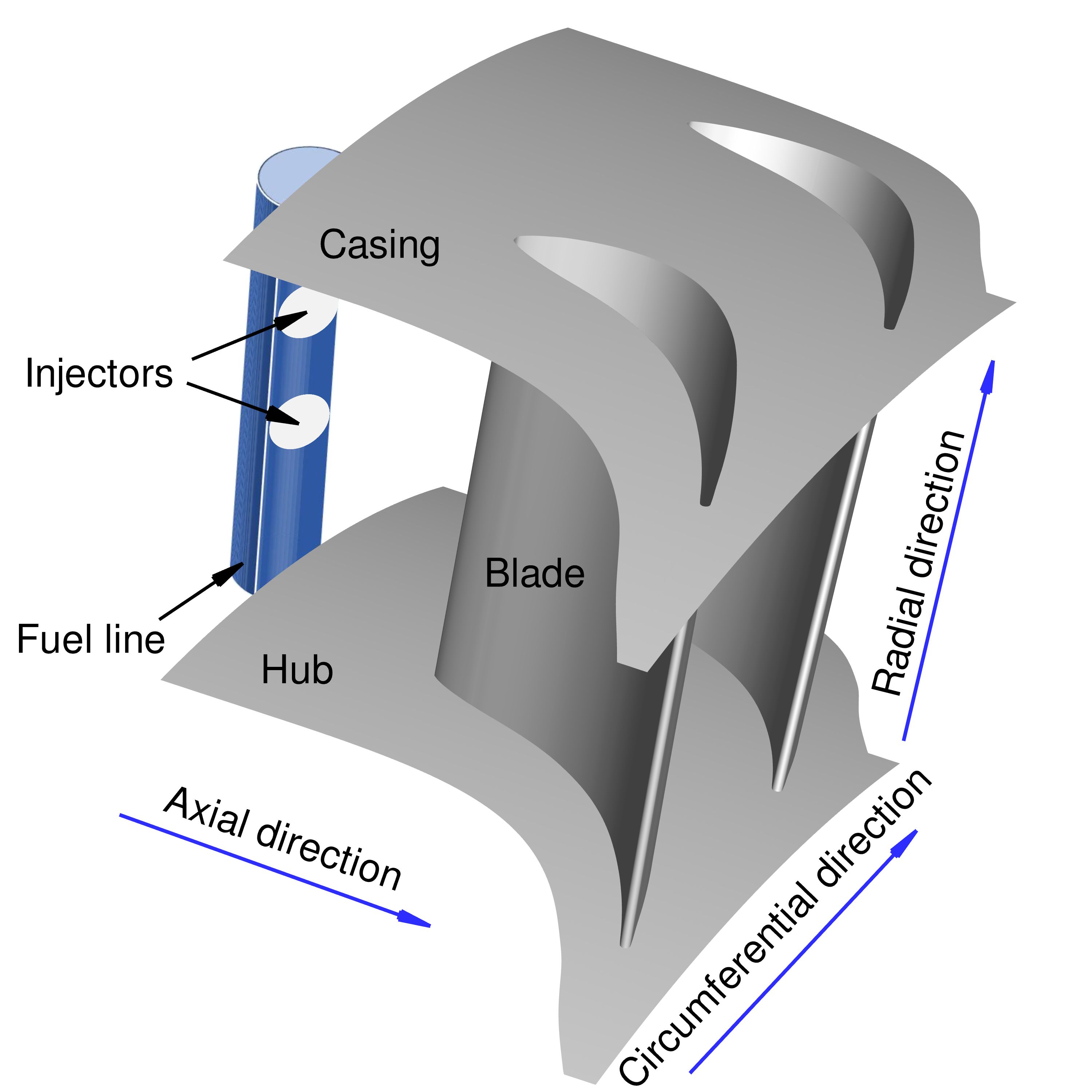}
    \caption{Sketch of fuel-injection structure at turbine inlet.}
    \label{fig:structure_configuration}
\end{figure}

This study considers a turbine-burner configuration in which cold fuel is injected parallelly into the hot vitiated air at the stator entrance. One possible structural design for this purpose is a simple strut configuration, as shown in Fig. \ref{fig:structure_configuration}. A straight fuel line with a semicircular cross section is installed along the spanwise direction ahead of the stator. The semicircular side of the strut faces the incoming flow while a series of circular fuel injectors are mounted on the flat back side. Fuel is delivered to the supply line in the strut through connection to either the hub, the casing, or both of the turbine stator. To enhance mixing of fuel and air downstream, a vortex generator is attached to each fuel injector to produce a swirling flow at the injector exit. The swirling fuel jet mixes with the hot vitiated air and then burns in the turbine passage on self ignition. 

\begin{figure}[htb!]
    \centering
    \includegraphics[width=0.48\linewidth]{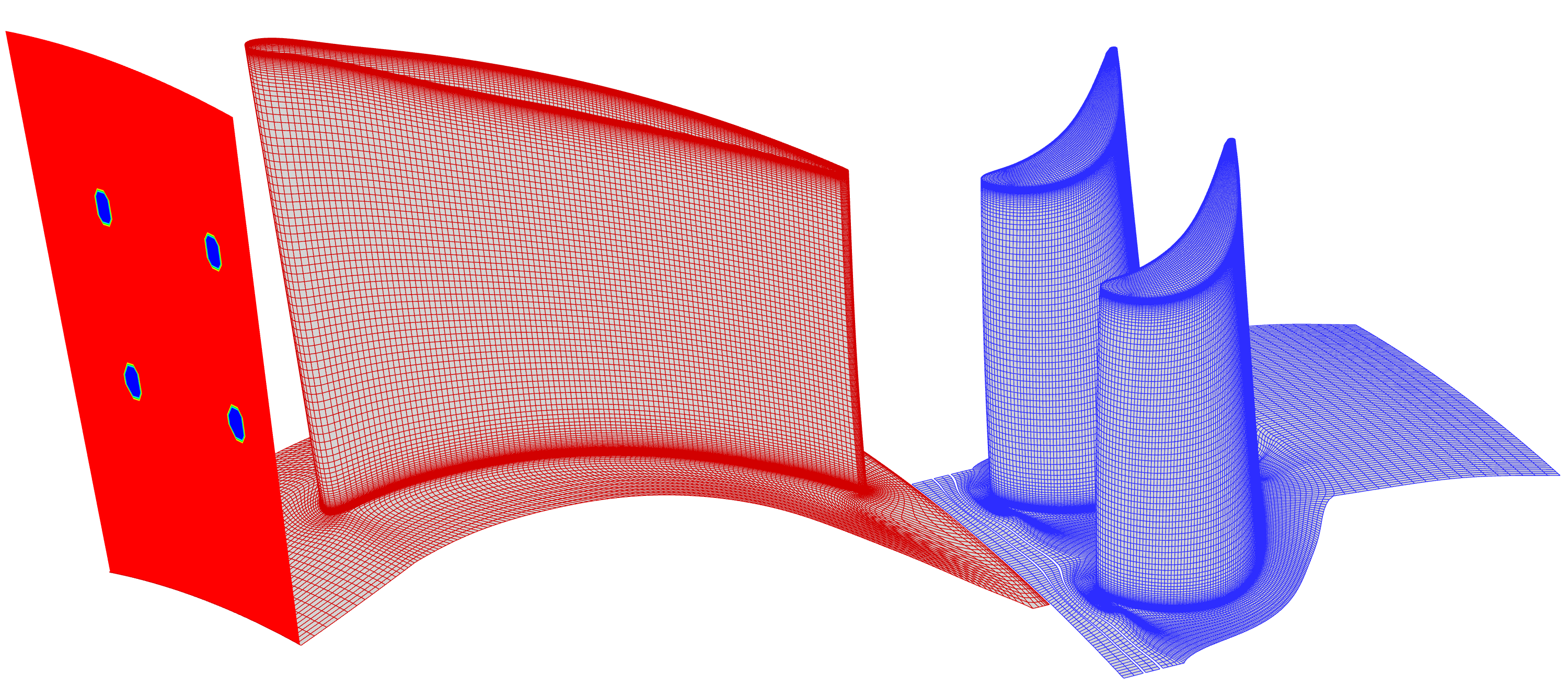}
    \caption{Computational domain and grids over the hub and blade surfaces. Every other grid point is shown.}
    \label{fig:grid_turbine_stage}
\end{figure}

Multi-block structured grids are generated within one stator and two rotor passages, as shown in Fig. \ref{fig:grid_turbine_stage}. 
The cell vertices on the interfaces between neighboring blocks are matched except at the block interfaces between the stator and rotor blade rows, where sliding patched grids and interpolation are used in order to account for the relative motion. The numbers of cells within one spanwise layer are 32,672 and 70,912 for the stator and rotor passages, respectively. There are 241 grid points along the spanwise direction for both the stator and the rotor, with points concentrated near the hub and casing. 33 grid points are distributed across the rotor tip clearance of $0.41 \, \mathrm{mm}$. This produces a computational domain with a total of over 26 million cells. The dimensionless distance $y^+$ of the first grid point away from wall is less than one. The grid aspect ratios satisfy $\Delta x / \Delta y < 80$ and $\Delta x / \Delta y < 50$ within the stator and rotor passages, respectively, while $\Delta z / \Delta y < 90$ in both the stator and rotor domains.

In the present work, we do not aim to study the details of the flow passing around the fuel-line/injector assembly. Therefore, the geometric details of the fuel line and injector assembly are omitted from the computation. Each fuel injector is modeled as a forced vortex with a lower temperature and the same pressure as the air. The flow inside and outside the fuel-line/injector assembly is not resolved. We focus on the aerodynamic and thermodynamic performances of the reacting flow through the turbine stage, on which neglect of the fuel-line/injector assembly has very minor impact because it is located at the entrance to the blade stator where the flow velocity is very low. Explicitly assessment of the fuel line and injector is an important future undertaking that will require refined geometric definition, extra meshing, and significant computation.

Turbomachinery blade rows are placed next to each other within very short distances. For computations of isolated blade row or a stage, the inlet and outlet boundaries are often extended upstream and downstream, respectively, by a distance comparable to the chord length of the blade to avoid undesirable reflection of perturbations by the imposed computational boundaries. In the present study, the outlet boundary is positioned 1.15 rotor chord length downstream of the rotor trailing edge following common practice. The inlet boundary, however, is placed at 0.36 stator chord length upstream of the stator leading edge, where the fuel injectors are installed. The fuel injectors are placed close to the blades to prevent premature combustion before the fuel reaches the turbine passage. Since we do not study the details of fuel injection and the flow passing around the fuel injector assembly in this paper, it is convenient to place the inlet boundary of the computational domain at the same location of the fuel injectors where discrete swirl injectors can be modeled as prescribed inlet flow conditions of the fuel. The slightly shorter distance  of the inlet boundary from the blades is expected to have minimum impact on the computational results, especially for comparison studies between the non-reacting and reacting flows since all computations are performed with the same computational domain. The inlet boundary will need to be placed further upstream in future studies when the impact of detailed flow over the fuel injector assembly must be considered.

\begin{figure}[htb!]
    \centering
    \includegraphics[width=0.49\linewidth]{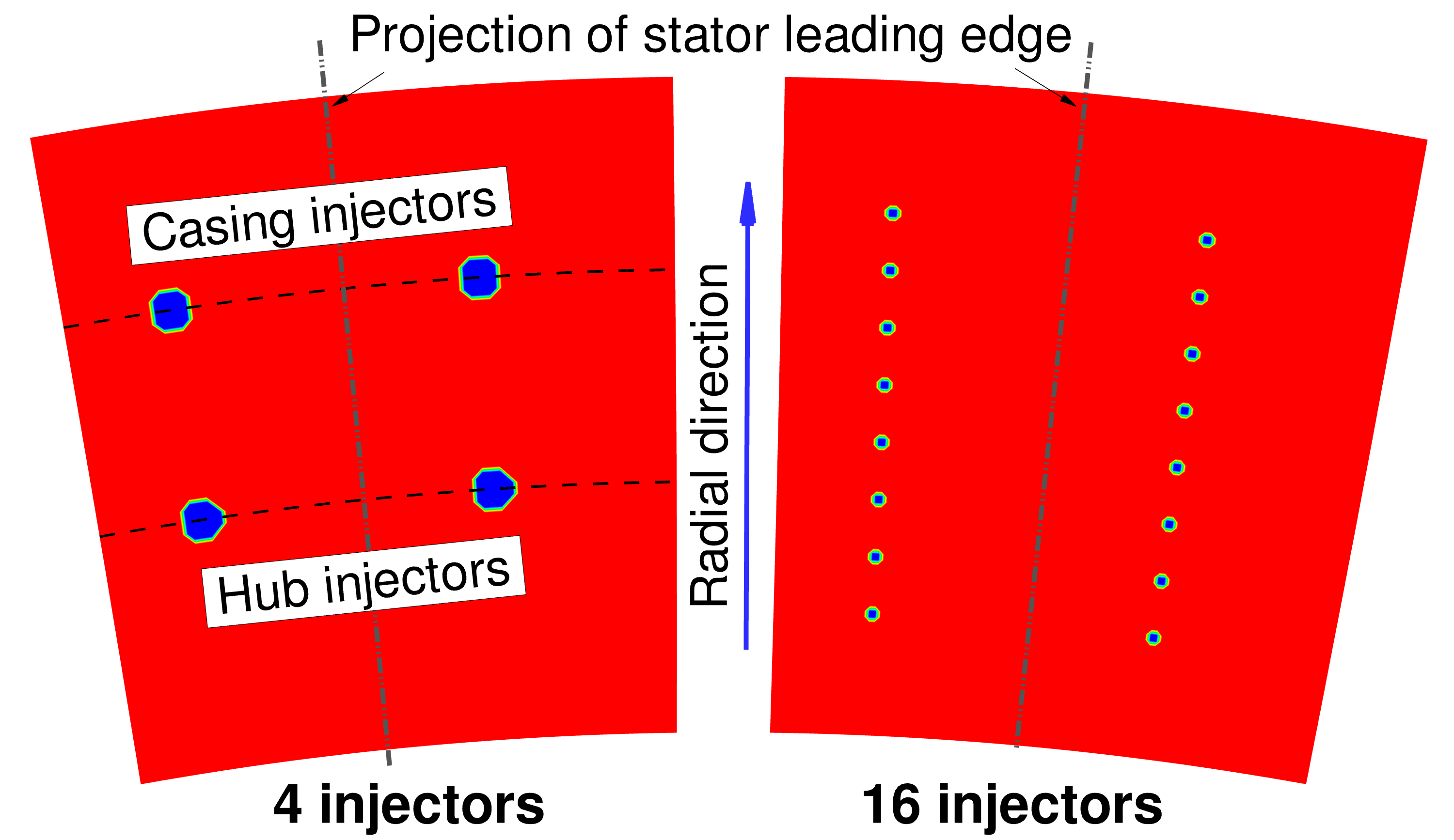}
    \caption{Sketch of fuel-injector configurations at turbine inlet.}
    \label{fig:sketch_4_16injectors}
\end{figure}

The flow downstream of the fuel-injector plane is simulated, as defined by the computational domain shown in Fig. \ref{fig:grid_turbine_stage}.
Four cases are considered:
\begin{description} 
   \item[Case 0N (0 injectors, nonreacting):] This is the baseline design representing the conventional turbine without combustion in it. The turbine inlet is supplied with uniform vitiated air, without fuel injection.
   \item[Case 4N (4 injectors, nonreacting):] At the turbine inlet, two fuel lines, each of which contains two fuel injectors with a radius of $2.0 \, \mathrm{mm}$, are placed in the middle of the blade passage, as shown on the left side of Fig. \ref{fig:sketch_4_16injectors} and in Fig. \ref{fig:grid_turbine_stage}. The two injectors near the hub (called the hub injectors) are on a circular arc of radius $335.8 \, \mathrm{mm}$ from the rotation center of the engine. The other two injectors (casing injectors) are at a radius of $355.8 \, \mathrm{mm}$. Nonreacting flow is simulated.
   \item[Case 4R (4 injectors, reacting):] The inlet fuel-injector configuration remains identical to that in case 4N; however, reacting flow is computed.
   \item[Case 16R (16 injectors, reacting):] Two fuel lines are placed at the same circumferential locations as in the 4-injector cases. However, each fuel line now contains eight small, equally spaced injectors, giving a total of sixteen, as shown on the right side of Fig. \ref{fig:sketch_4_16injectors}. This case is designed to produce a more radially uniform fuel injection than case 4R, while maintaining the same total fuel mass flow rate. Reacting flow is computed.
\end{description}

The fuel injectors in cases 4N, 4R, and 16R are positioned away from the hub, casing, and stator blade surfaces (as indicated by the projection of stator leading edge in Fig. \ref{fig:sketch_4_16injectors}), to keep the combustion centered in the stator passage and thus to minimize the additional thermal loading on the stator walls resulting from the combustion.

\begin{figure}[htb!]
    \centering
    \begin{subfigure}[b]{0.49\linewidth}
        \centering
        \includegraphics[width=1\linewidth]{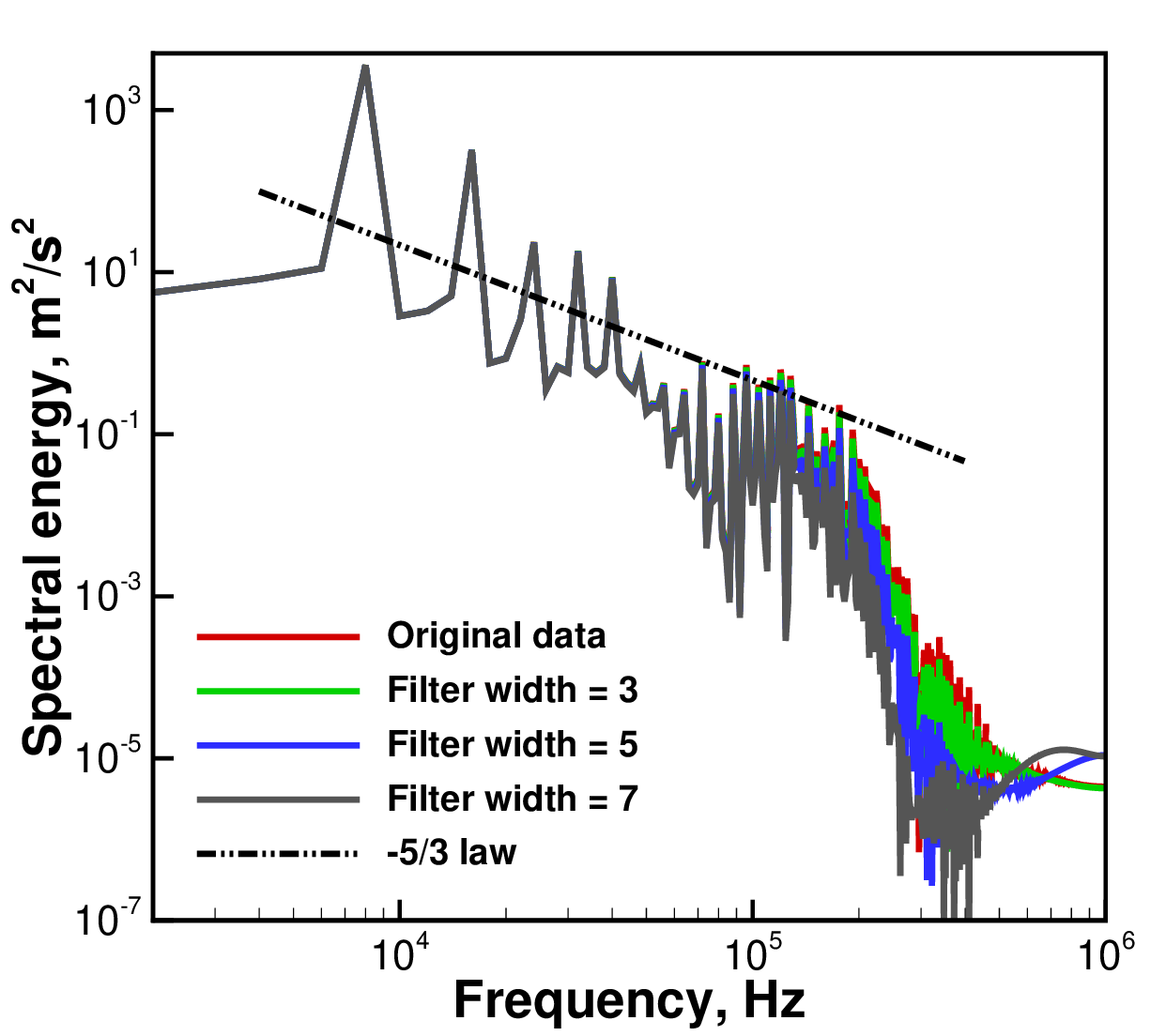}
        \caption{Temporal spectral energy}
        \label{fig:energy_spectrum_time}
    \end{subfigure}
    \begin{subfigure}[b]{0.49\linewidth}
        \centering
        \includegraphics[width=1\linewidth]{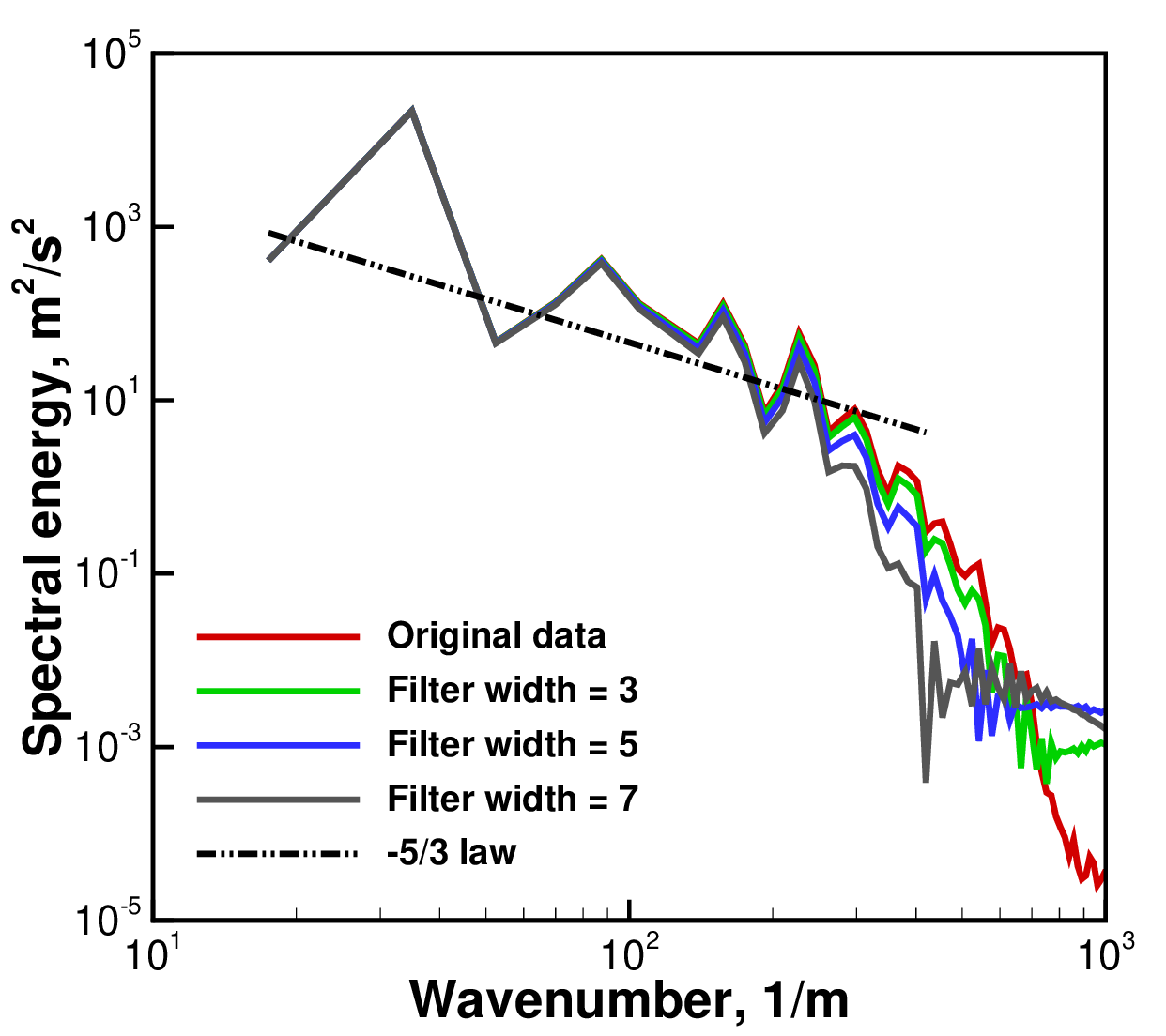}
        \caption{Circumferential spectral energy}
        \label{fig:energy_spectrum_pitch}
    \end{subfigure}
    \caption{Spectral energy of axial velocity  near the rotor-inlet hub in case 4R.}
    \label{fig:energy_spectrum}
\end{figure}

The temporal spectral energy at a point near the rotor-inlet hub in case 4R is obtained by applying a Fourier transform to the time series of axial velocity perturbations, as shown in Fig. \ref{fig:energy_spectrum_time}. The figure also includes results obtained after temporally filtering the time series using different filter widths.
Figure \ref{fig:energy_spectrum_pitch} shows the spatial spectral energy of the axial velocity perturbations along the circumferential direction of the rotor-inlet hub.  
Both the temporal and the spatial energy spectra exhibit the -5/3 scaling law within their respective inertial frequency and wavenumber ranges. The discrete energy peaks observed in Fig. \ref{fig:energy_spectrum_time} correspond to the blade passing frequency of the stator and its higher-order harmonics.
In addition, as the filter width increases, the energy associated with high frequencies and high wavenumbers decreases, whereas the inertial-range spectra remain unchanged.
These results indicate that the LES computation based on the current mesh and time-step captures most of the turbulent fluctuation features.
With a finer grid resolution, a broader inertial range would be expected to emerge.

\subsection{Boundary and Initial Conditions}
There are five types of boundaries in the computational domain: inlet, outlet, wall, periodic boundaries, and stator-rotor interface.
The blade surfaces, hub, and casing of the turbine are all specified as no-slip adiabatic walls with zero normal pressure and mass fraction gradients. The blade surface and hub of the rotor are specified to rotate with the rotation speed of the turbine while the other walls are set as stationary.
The circumferential boundaries in the computational domain are treated as rotating periodic boundaries.
The sliding mesh technique is used to treat the interface between the stator and rotor. The volume-weighted interpolation method developed for multi-row turbomachines by Zhu et al. \cite{zhu2017numerical, zhu2018flow, liu2025computational} is applied to exchange flow variables on the interface.

At the turbine inlet, vitiated air is used to simulate the exhaust gas from the upstream primary combustor of a gas-turbine engine. The pressure and temperature for air are $30 \, \mathrm{bar}$ and $1645 \, \mathrm{K}$, respectively. It is estimated that, to reach a turbine inlet temperature of $1645 \, \mathrm{K}$, a fuel-air mass ratio of 0.03 is needed in the primary combustor. As a result, the vitiated air at the turbine inlet consists of $73.77\%$ $\rm{N_2}$, $11.01\%$ $\rm{O_2}$, $8.04\%$ $\rm{CO_2}$, and $7.18\%$ $\rm{H_2O}$ by mass fraction. The air flows into the domain at $80 \, \mathrm{m/s}$ along the axial direction. 
Fuel is injected at the same pressure as the air but at a lower temperature of $395 \, \mathrm{K}$. The axial velocity of fuel is $15 \, \rm{m/s}$, which produces a mass ratio of fuel to air of $0.008$ at the inlet.
In cases 4N and 4R, the swirling velocity inside each injector is specified by a forced vortex aligned with the positive axial direction, with a maximum velocity of $50 \, \rm{m/s}$ at the injector edge. In contrast, no swirling is imposed in case 16R due to the smaller injector radius.
The total pressure, total temperature, and flow angles, computed from the above pressure, temperature, and velocity, are specified as inlet boundary conditions along with the mass fraction of each species. A cell at the inlet is treated as part of the fuel region if its centroid lies within the fuel injector; otherwise, it is designated as part of the air region.
In addition, the turbulence intensity and the ratio of turbulent to molecular viscosities at the inlet are set to $10\%$ and 10.0 at the whole inlet, respectively. They are used to synthesize realistic turbulent fluctuations at the inlet by the random flow generation (RFG) method proposed by Smirnov et al. \cite{smirnov2001random}.

Since the flow at the outlet is subsonic with no reverse flow, one flow variable at the outlet plane must be specified to maintain the well-posedness of the problem. In the annular turbine, spanwise variation of static pressure is assigned by the simple radial equilibrium equation with the pressure at the hub being fixed \cite{liu2025computational}. The hub pressure is set to 0.4 times the total pressure at the inlet. The streamwise gradients of other variables are set to zero. 

Carefully constructed initial conditions are used to quickly establish the operating conditions. At the initial stage, the entire domain is filled with vitiated air with a temperature of $1645\, \mathrm{K}$ and a pressure of $30 \, \mathrm{bar}$. A steady nonreacting computation, using local time steps determined by the flow within each grid cell, is first performed. It is then transferred to a steady reacting computation to quickly establish the reacting flow in the domain. Once a converged solution is achieved, the simulation is switched to the unsteady mode.
During the unsteady computation, the physical time step is determined as the minimum of the time steps over the entire domain, subject to a Courant–Friedrichs–Lewy (CFL) number of 0.8, and is typically on the order of $10^{-6} \, \mathrm{ms}$.
The time interval for outputting the flow field is $5 \times 10^{-4} \, \mathrm{ms}$, resulting in 125 flow-field outputs per blade passing period (BPP) of the rotor.
The BPP, defined as the time for the rotor blade to rotate by one pitch angle of its passage, is $0.0625 \, \mathrm{ms}$ in this case.

\section{Results and Discussions} \label{sec:results_discussions}

\subsection{Reacting Flow Field} \label{sec:reacting_flow}
\begin{figure}[b!]
    \centering
    \begin{subfigure}[b]{0.49\linewidth}
        \centering
        \includegraphics[width=1\linewidth]{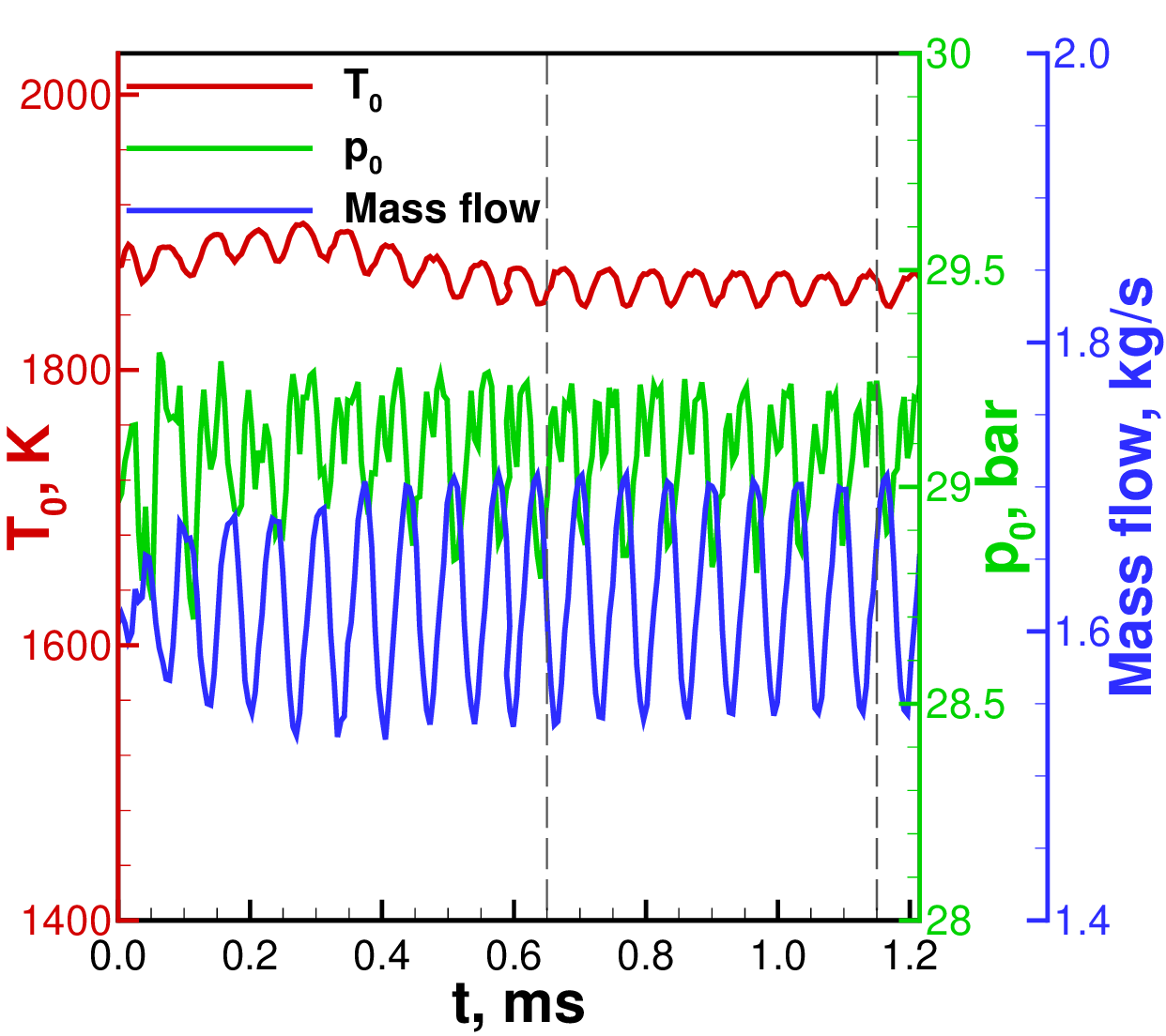}
        \caption{Stator outlet}
    \end{subfigure}
    \begin{subfigure}[b]{0.49\linewidth}
        \centering
        \includegraphics[width=1\linewidth]{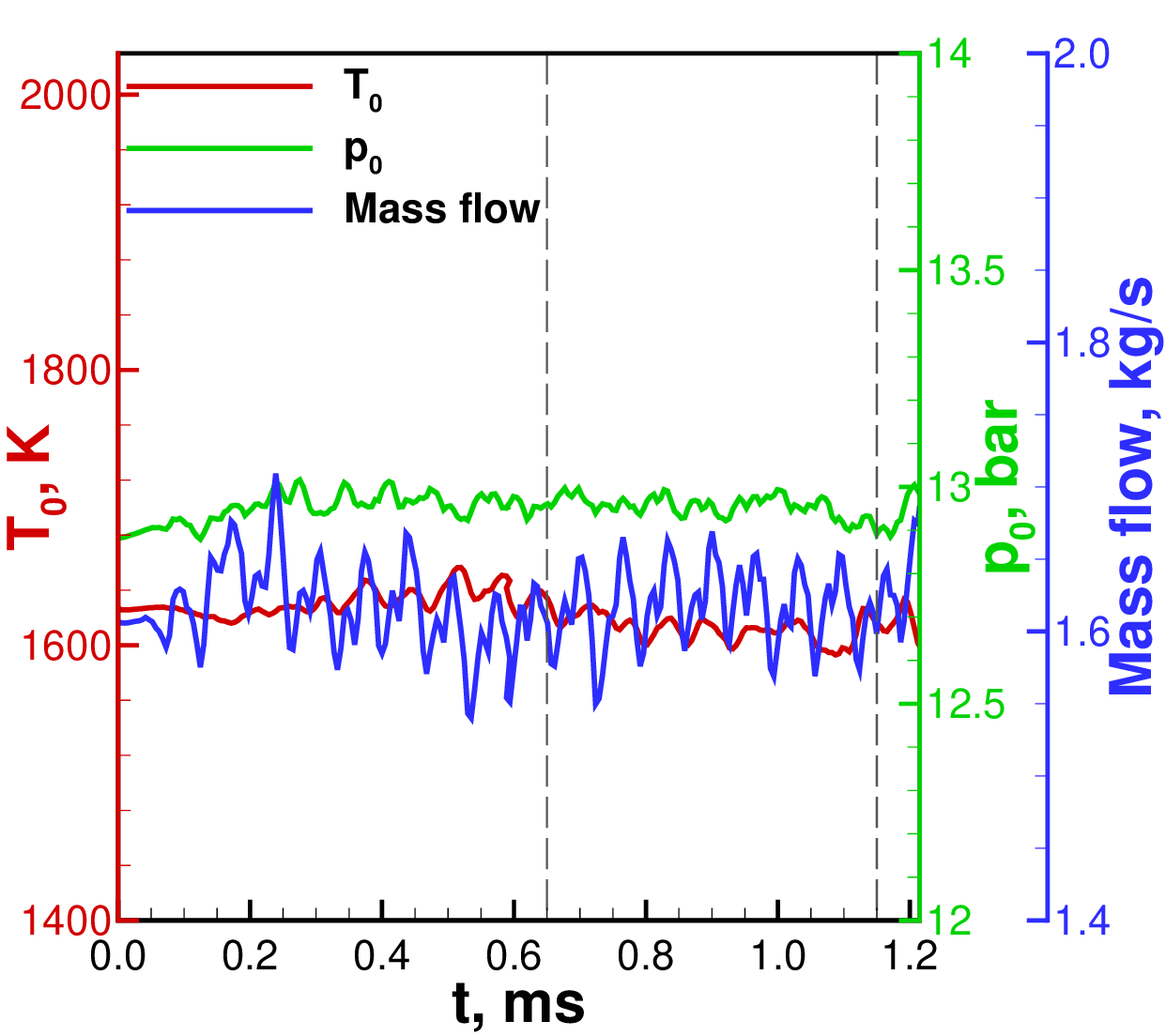}
        \caption{Rotor outlet}
    \end{subfigure}
    \caption{Time history of flow quantities averaged over the stator and rotor outlets in case 4R.}
    \label{fig:P0_T0_mass-time}
\end{figure}

The unsteady reacting flow field by LES is analyzed in this section. Since the flow fields of cases 4R and 16R exhibit very similar behaviors, only the results of case 4R are presented here without loss of generality. 
Figure \ref{fig:P0_T0_mass-time} shows the time history of the total temperature, total pressure, and mass flow rate averaged over the stator and rotor outlets. Since a precursory steady solution is used as the initial condition for the unsteady simulation, a quasi-periodic state is quickly established in the turbine stage after approximately $0.6 \, \rm{ms}$. 
In the quasi-periodic regime, the peak-to-peak oscillation amplitudes of total temperature, total pressure, and mass flow rate at the stator outlet are $23 \, \rm{K}$, $0.4 \, \rm{bar}$, and $0.16 \, \rm{kg/s}$, respectively. They become $15 \, \rm{K}$, $0.08 \, \rm{bar}$, and $0.08 \, \rm{kg/s}$ at the rotor outlet. 

The flow exhibits strong oscillations of total temperature, total pressure, and mass flow rate at the interface between the stator and rotor with a clear large-scale frequency that corresponds to the rotor BPP, $0.0625 \, \mathrm{ms}$. This is because of the strong stator-rotor interaction and also the fact that most chemical reaction occurs within the upstream stator passage, as discussed later. The oscillations at the rotor outlet are significantly reduced because there is little chemical reaction in the rotor passage and no downstream rotor-stator interaction at the rotor outlet is considered. A constant static pressure condition is imposed there.
The flow-through time across the turbine stage is estimated to be $0.266\, \rm{ms}$, close to 4 BPPs. The flow fields between $0.65 \, \rm{ms}$ and $1.15 \, \rm{ms}$, corresponding to 8 BPPs or approximately two times the flow-through time, are selected for analyzing the aerodynamic performance of the turbine stage hereafter, unless otherwise stated.

\begin{figure}[htb!]
    \centering
    \includegraphics[width=0.49\linewidth]{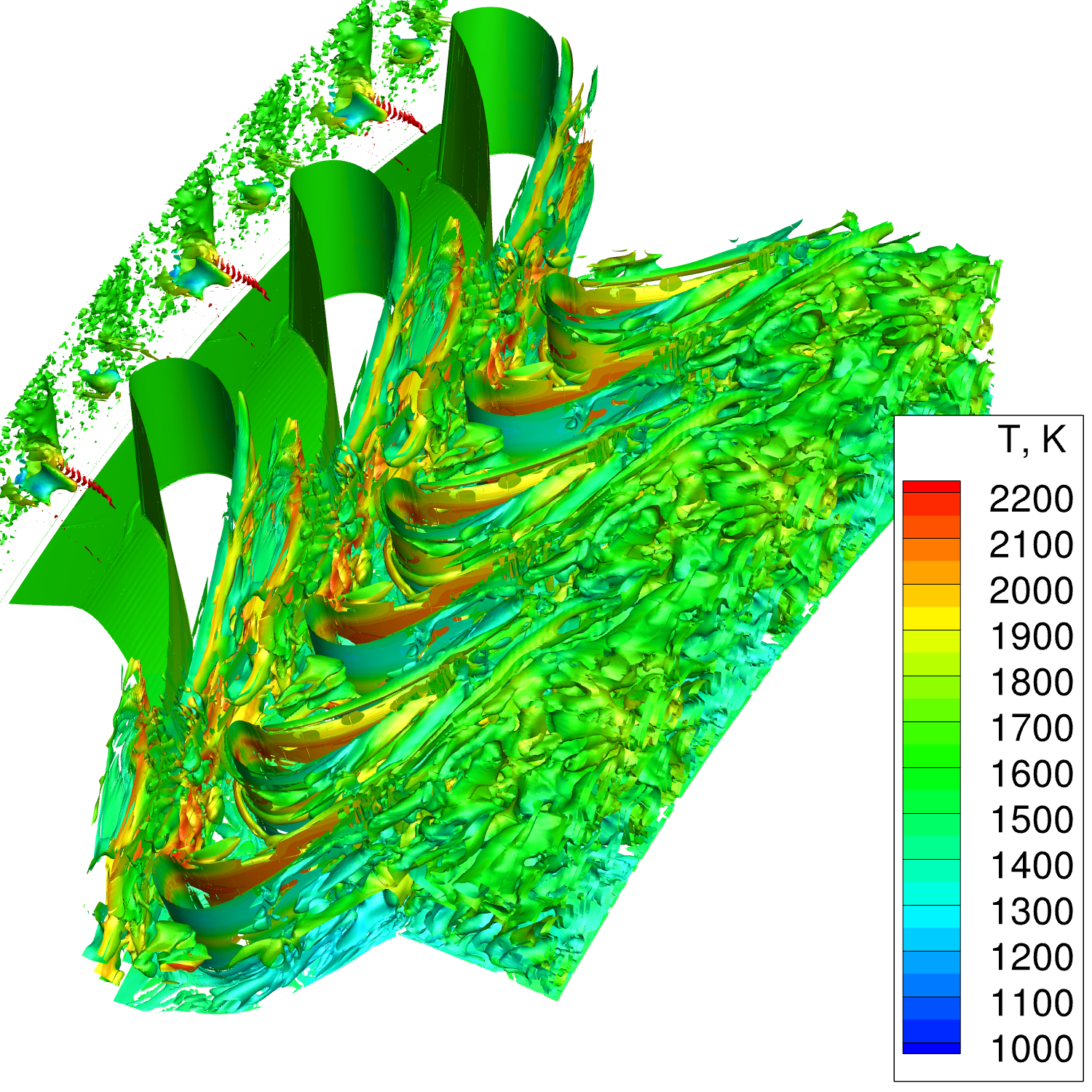}
    \caption{Instantaneous vortex structures identified by the $\mathbf{\lambda_2}$ criterion and colored by temperature in the hub half of the turbine stage for case 4R.}
    \label{fig:vortex_reacting_4injectors}
\end{figure}

Figure \ref{fig:vortex_reacting_4injectors} shows the instantaneous vortex structures identified by the $\lambda_2$ criterion \cite{jeong1995identification} and colored by temperature in the hub half of the turbine stage in case 4R. 
Shear layers form on the tubular-like fuel-air interfaces at the turbine inlet and gradually become unstable in the latter half of the stator passage. After mixing with the vortices shed from the stator trailing edge, the eddies are split into two streams by the rotor leading edge, reoriented in the transverse direction, and transported downstream along both the pressure and suction surfaces of the rotor blade. 
The temperature on the fuel-air interface is higher than the inlet air temperature due to chemical reactions.
As the flow develops, large-scale coherent structures are generated in the rotor passage due to stator-rotor interactions, along with more small-scale eddies.
The temperature maintains at similar levels along the streamwise direction in the rotor passage, unlike in the nonreacting cases where it decreases. This difference will be further examined in Sec. \ref{sec:aerodynamic_performance}.

\begin{figure}[htb!]
    \centering
    \begin{subfigure}[b]{0.33\linewidth}
        \centering
        \includegraphics[width=1\linewidth]{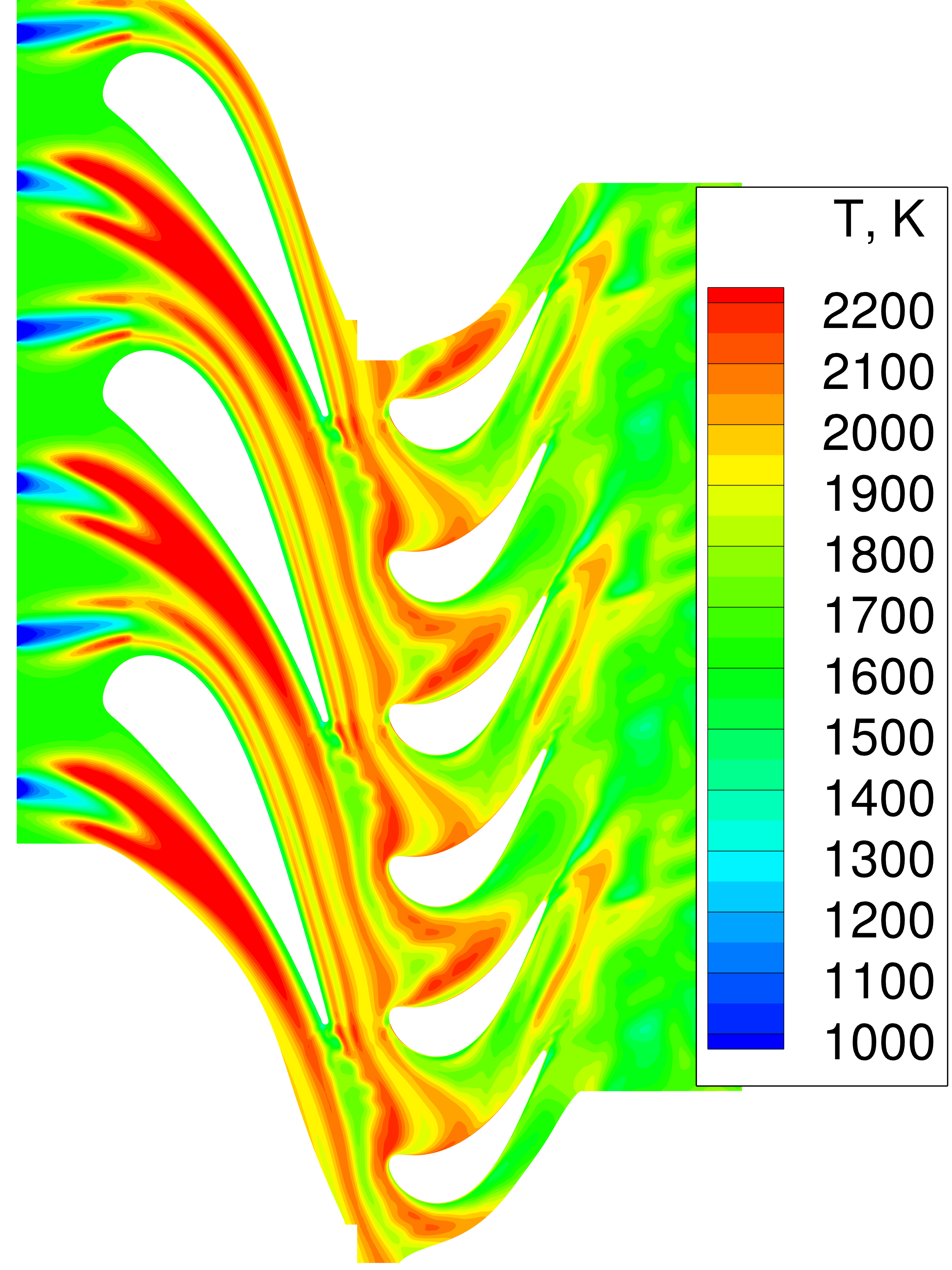}
        \caption{Temperature}
        \label{fig:contour_instant_radial_slice_T}
    \end{subfigure}
    \begin{subfigure}[b]{0.33\linewidth}
        \centering
        \includegraphics[width=1\linewidth]{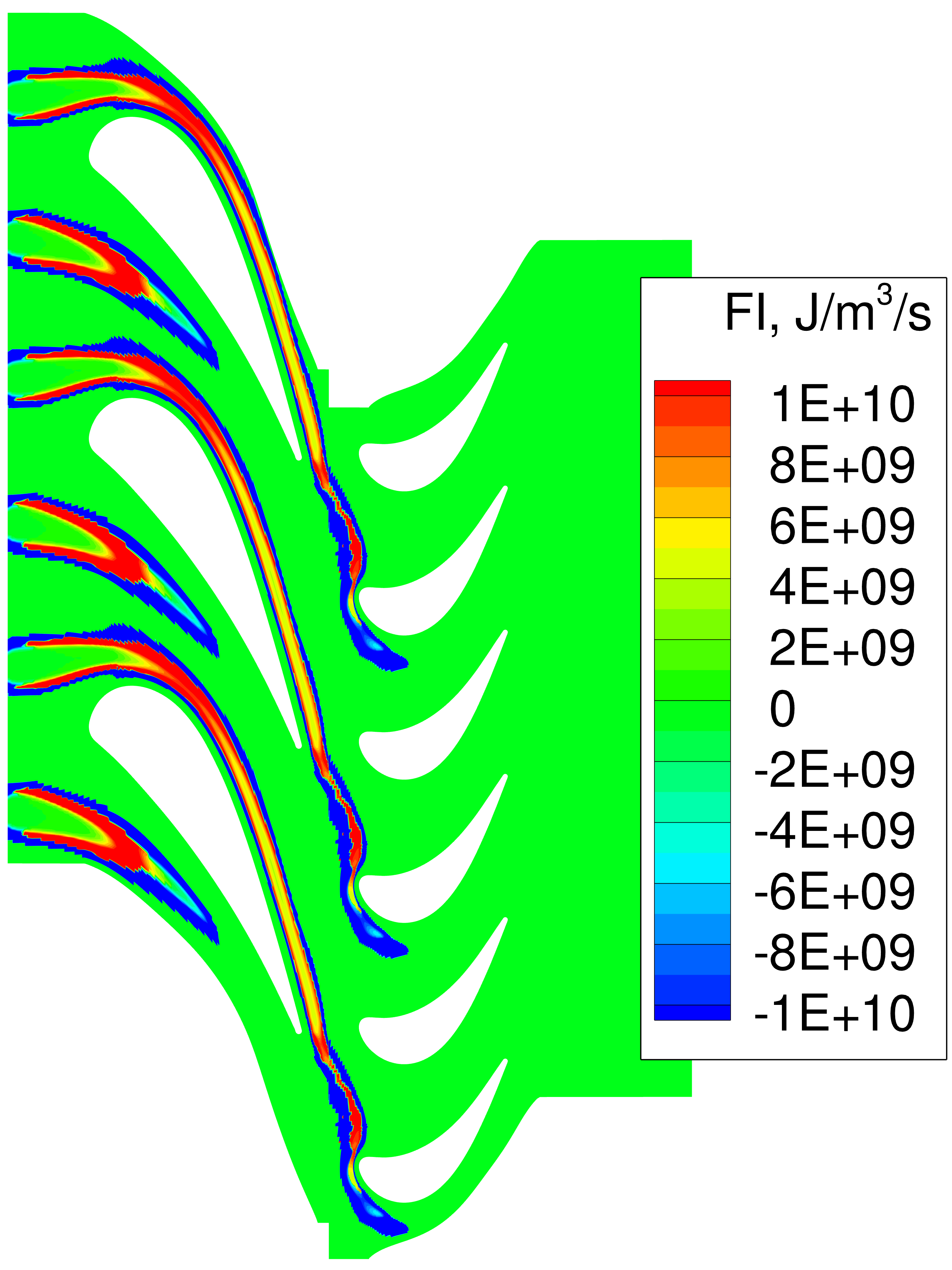}
        \caption{Flame index}
        \label{fig:contour_instant_radial_slice_FI}
    \end{subfigure}
    \caption{Instantaneous contours on the B2B plane passing through hub-injector centers in case 4R.}
    \label{fig:contour_instant_radial_slice}
\end{figure}

Figure \ref{fig:contour_instant_radial_slice} shows the instantaneous contours of temperature and flame index on the B2B plane that passes through the hub-injector centers. Here, a B2B plane is defined as a surface of revolution with a specific radius. Only the hub-injector B2B plane is shown here for brevity. The flow in the casing-injector B2B plane is similar. The computational domain is duplicated twice in the circumferential direction to better illustrate the variations across turbine passages.
The flame index (FI) is defined as the heat release rate per unit volume masked by the sign of the dot product of the gradients of fuel and oxidizer mass fractions, i.e.,
\begin{equation} \label{eq:flame_index}
    \mathrm{FI} = \mathrm{sign} \left( \nabla Y_\mathrm{CH_4} \cdot \nabla Y_\mathrm{O_2} \right) \dot Q
\end{equation}
where the heat release rate is given by Eq. (\ref{eq:hrr}). If the flame index is positive, fuel and oxidizer diffuse in the same direction, indicating a premixed flame; otherwise, it indicates a diffusion flame. Using the mass-fraction gradients normal to the velocity vector in Eq. (\ref{eq:flame_index}) yields the same flame-index distributions for the present cases.

Two flames (tubular-like in three dimensions due to the discrete round fuel injectors) are established in each stator passage. They follow the main flow through the blade passage.
One of them goes through the stator passage near the suction surface (suction-surface flame) and the other one near the pressure surface (pressure-surface flame).
Once ignited, both strength and thickness of the two flames increase rapidly in the low-speed region near the stator entrance.
As a result, the radii of the two flames are significantly larger than those of fuel injectors at the inlet. 
The intensity and growth of the flame, however, attenuate starting shortly downstream due to decreasing pressure and temperature and the stretching of the flames by the fast accelerating flow in the stator.
As discussed in the previous work \cite{zhu2024numerical}, a favorable pressure gradient suppresses chemical reaction.
The flow accelerates much faster on the suction surface than on the pressure surface, resulting in a faster decrease of the pressure and therefore much thinner and weaker flame near the suction surface, as shown by the temperature contours in Fig. \ref{fig:contour_instant_radial_slice_T}. The high and near-constant pressure on the pressure surface and near the leading edge of the stator blade gives rise to the intense combustion in that region, which is evidenced by the wider and higher temperature and flame index zones shown in Fig. \ref{fig:contour_instant_radial_slice_T} and \ref{fig:contour_instant_radial_slice_FI}, respectively. However, the high flame index by the pressure side dies out quickly before it reaches the blade trailing edge while the flame by the suction side continues beyond the blade trailing edge despite being at low intensities. The fast reaction in the pressure-surface flame has consumed all of its fuel in the middle of the passage. The high-temperature zone continues beyond the reaction zone due to convection but its size and level reduce because of flow acceleration.
The flame by the suction surface continues through the passage and is enhanced by the increased mixing in the wake of the blade and the stator-rotor interaction, as shown in both Figs. \ref{fig:contour_instant_radial_slice_T} and \ref{fig:contour_instant_radial_slice_FI}.
These flame variations in the stator passage are similar to the 2D RANS results in a turbine cascade reported in the previous work \cite{zhu2024numerical}. 

On impact with the rotor blade, the hot streaks and the remaining flame are each split into two streams by the rotor blades and convected downstream.
Due to rotor-stator interactions, mixing and consequently combustion are enhanced in the rotor passage, as indicated by the increased flame index of the continued suction-surface flame. The flame quenches shortly downstream of the rotor leading edge due to fuel depletion.
In each rotor passage, the streak impinges on the pressure surface and is then convected towards the suction surface of the adjacent blade due to the transverse pressure gradient. The hot streak originating from the stator suction-surface flame persists longer within the rotor passage compared with that from the pressure-surface flame, because the former undergoes additional combustion in the stator wake and near the rotor leading edge.
This may lead to a high temperature level on the rotor blade, thus increasing the heat load on it.

The variations of flame index are consistent with those of temperature. Fuel is ignited as a diffusion flame immediately at the turbine inlet, as indicated by the negative flame index. In each round reacting region, the inner side is occupied by premixed flames, as evidenced by the positive flame index. This is attributed to the pre-swirling of fuel at the turbine inlet. In contrast, the outer side, directly interacting with the surrounding air, is dominated by diffusion flames.
After the stator trailing edge, the shear layer between fuel and air becomes unstable, which enhances the mixing of unburnt fuel and air. As a result, a strong premixed flame is generated in the stator wake and further enlarged near the rotor leading edge. Nevertheless, the combustion in the rotor is still dominated by diffusion flames, similar to the situations in the stator.

\begin{figure}[htb!]
    \centering
    \begin{subfigure}[b]{0.49\linewidth}
        \centering
        \includegraphics[width=1\linewidth]{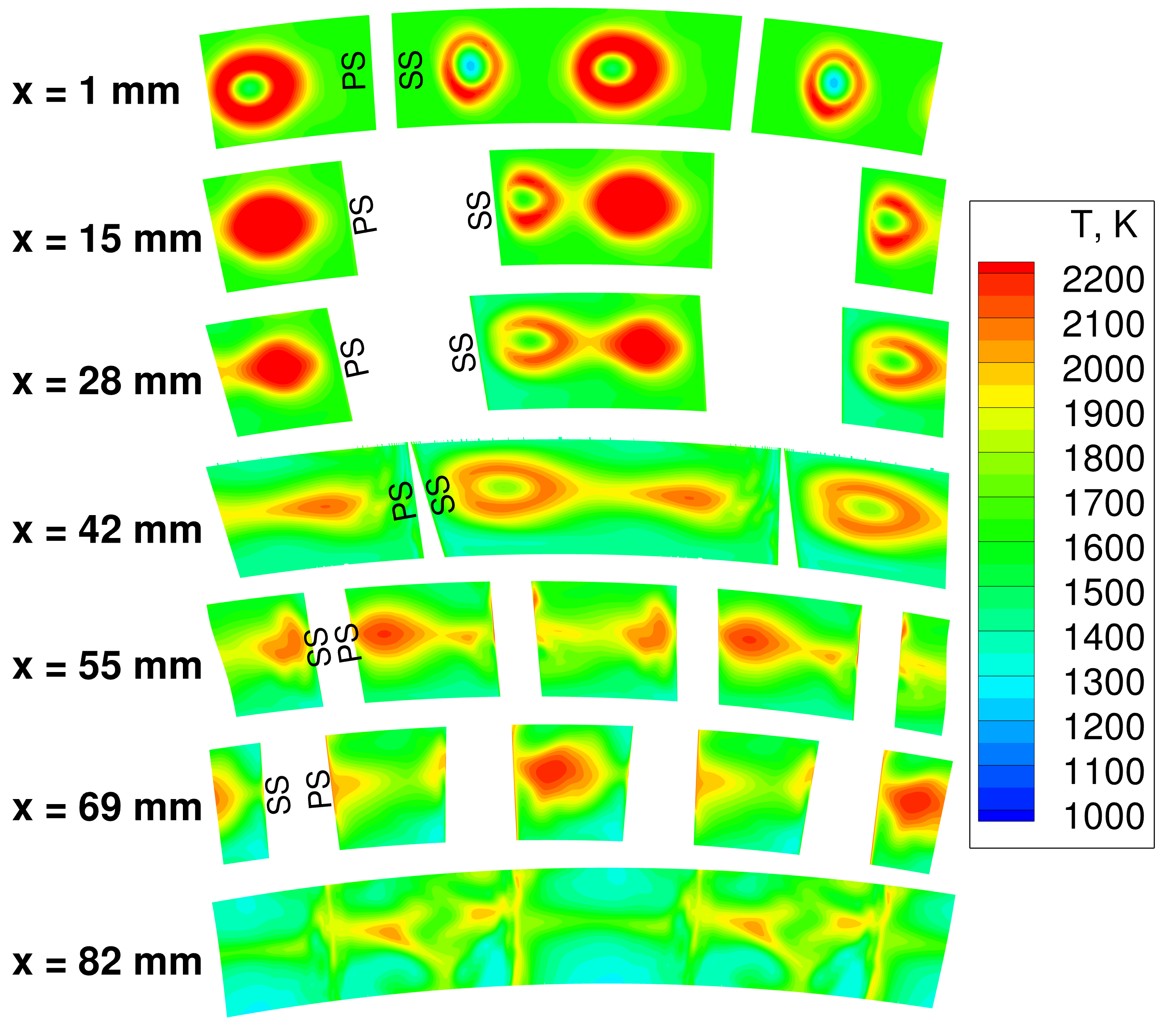}
        \caption{Temperature}
        \label{fig:contour_instant_stream_slice_T}
    \end{subfigure}
    \begin{subfigure}[b]{0.49\linewidth}
        \centering
        \includegraphics[width=1\linewidth]{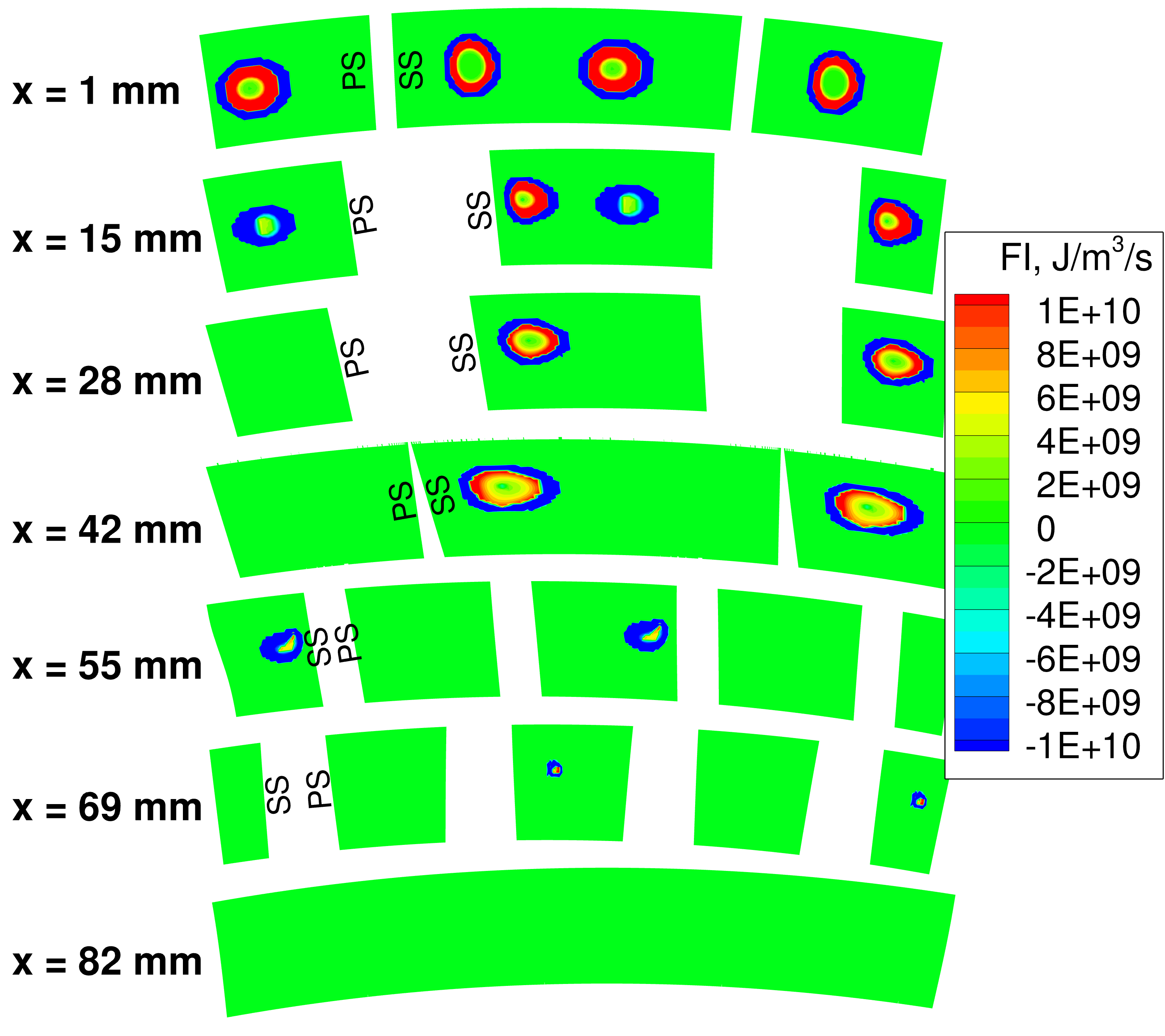}
        \caption{Flame index}
        \label{fig:contour_instant_stream_slice_FI}
    \end{subfigure}
    \caption{Instantaneous contours on axial slices near hub injectors in case 4R. The stator leading edge is at x = 0. SS: suction surface; PS: pressure surface.}
    \label{fig:contour_instant_stream_slice}
\end{figure}

To examine the evolution of flames along the radial direction, the instantaneous contours of temperature and flame index on flow cross sections at seven axial locations are shown in Fig. \ref{fig:contour_instant_stream_slice}. The first four slices are located in the stator passage while the left three ones are in the rotor passage. Each slice is centered on the hub-injector centers and extends three injector diameters in the radial direction. 
Near the stator leading edge, the flames begin stretching along the circumferential direction due to the convection of nonzero and non-uniform transverse velocity induced by the stator blades. In contrast, the flame develops slowly along the radial direction due to the almost zero radial velocity. This is also the reason why the reacting regions, as shown by the flame index, are larger in the circumferential direction than those in the radial direction. The circular hot streak near the pressure surface is almost stretched into a circumferential line at the stator trailing edge. 
In the stator wake, unsteady stator-rotor interactions and turbulent fluctuations keep stretching these flames and hot streaks, especially in the circumferential direction. As a result, near the rotor leading edge, all flames and hot streaks have met and merged into a nearly circumferential-aligned line, with small radial extension.

In the rotor passage, the flames and hot streaks keep extending in both directions and distorting due to large-scale unsteady motions and small-scale turbulent fluctuations. Consistent with Fig. \ref{fig:contour_instant_radial_slice}, flames are transported from the pressure surface to the suction surface in the rotor passage.
After the mid-chord of the rotor, the flames are extended to over six times the injector diameters in the radial direction, occupying the whole slice region. Recall that the distance between the hub injectors and casing injectors is five times the injector diameters. This means that the hub-injector flames have already met with the casing-injector flames.

\begin{figure}[b!]
    \centering
    \begin{subfigure}[b]{0.33\linewidth}
        \centering
        \includegraphics[width=1\linewidth]{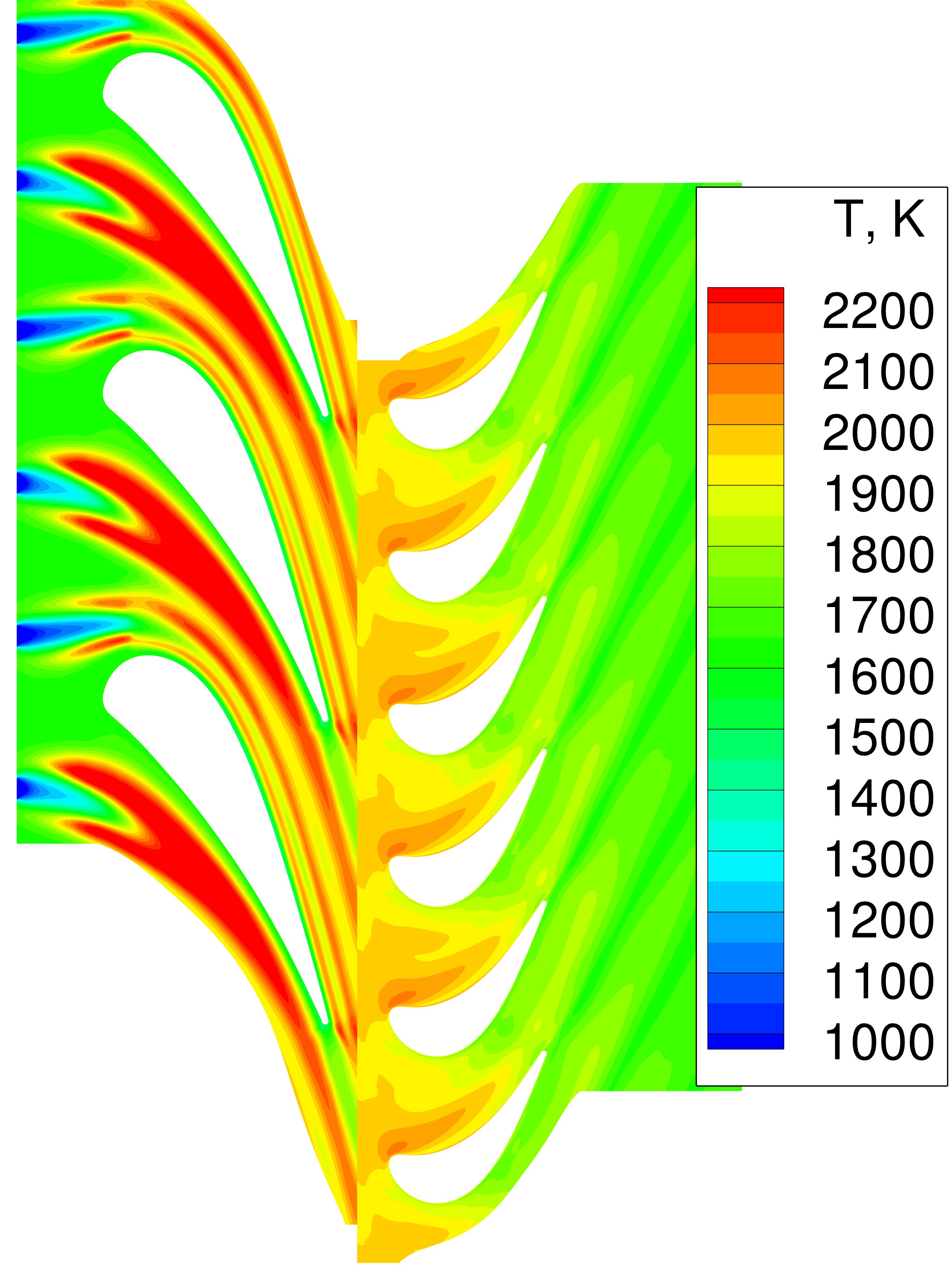}
        \caption{Temperature}
        \label{fig:contour_TA_radial_slice_T}
    \end{subfigure}
    \begin{subfigure}[b]{0.33\linewidth}
        \centering
        \includegraphics[width=1\linewidth]{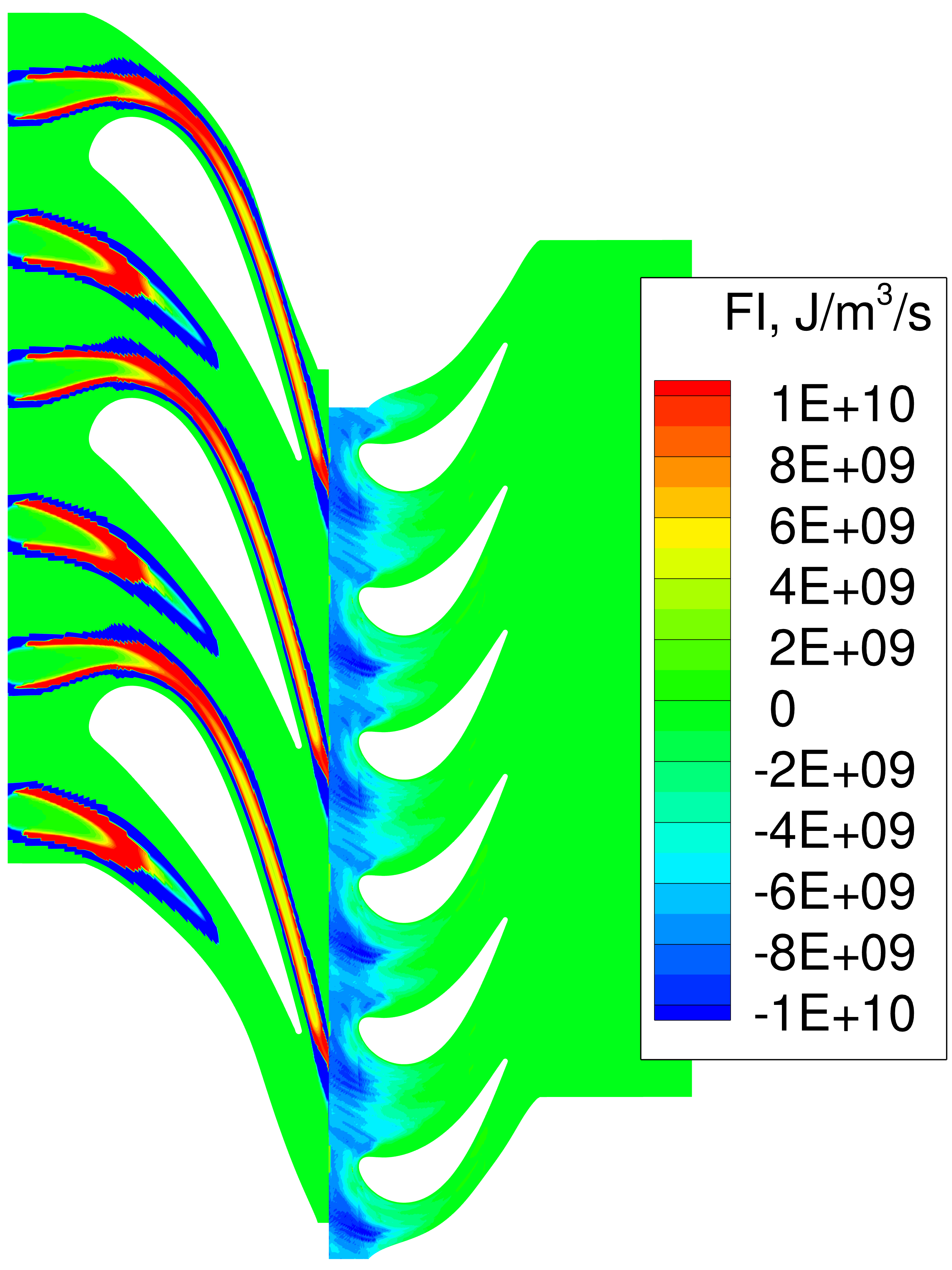}
        \caption{Flame index}
        \label{fig:contour_TA_radial_slice_FI}
    \end{subfigure}
    \caption{Time-averaged contours on the B2B plane passing through hub injector centers in case 4R.}
    \label{fig:contour_TA_radial_slice}
\end{figure}

Figure \ref{fig:contour_TA_radial_slice} shows the time-averaged contours of temperature and flame index on the same B2B plane as that in Fig. \ref{fig:contour_instant_radial_slice}. In the stator, both the time-averaged temperature and flame index are almost the same as those at a time instant in Fig. \ref{fig:contour_instant_radial_slice} because the upstream influence of the spinning rotor is weak.
However, differences between the averaged and instant flows appear in the stator wake and in the rotor passage due to the stator-rotor interaction and the downstream influence of the stator wake. 
In the rotor, the time-averaged temperature gradually decreases along the streamwise direction not only because the rotor blade extracts work, converting thermal energy into mechanical energy, but also because flow acceleration converts thermal energy into kinetic energy.
Near the rotor leading edge, the entire rotor pitch exhibits strong chemical reactions, as indicated by the high negative flame index. This is consistent with the high temperature observed in Fig. \ref{fig:contour_TA_radial_slice_T}.
The time-averaged temperature near the pressure surface is higher than that near the suction surface since the flames and hot streaks from the stator impinge on the rotor pressure surface, as shown by the instantaneous temperature in Fig. \ref{fig:contour_instant_radial_slice_T}.

\subsection{Aerodynamic Performance} \label{sec:aerodynamic_performance} 
In this section, the influence of combustion on the aerodynamic performance of the turbine stage is analyzed by comparing the reacting cases with the nonreacting cases. All flow quantities are computed based on the time-averaged flow field of LES computations.

Integrating the time-averaged energy equation (\ref{eq:energy}) over the computational domain $\Omega$ gives
\begin{equation} \label{eq:energy_balance}
    (\dot{m} \bar H)_{\mathrm{out}} - (\dot{m} \bar H)_{\mathrm{in}} + P_t + \dot \varepsilon_f = \dot Q_t 
\end{equation}
where $(\dot{m} \bar H)$ is the flux of total enthalpy across a surface. $(\cdot)_{\mathrm{in}}$ and $(\cdot)_{\mathrm{out}}$ represent the inlet and outlet of the turbine stage, respectively.
$P_t$ is the turbine power extracted from the flow by the rotor blade, computed by
\begin{equation} \label{eq:work_power}
    P_t = \mathbf{\Omega} \cdot \mathbf{M} = \mathbf{\Omega} \cdot \int_{\mathrm{rotor \ blade}} \mathbf{r} \times (p\mathbf{I} - \boldsymbol{\tau}) \cdot \mathrm{d}\mathbf{S}
\end{equation}
where $\mathbf{S}$ is the surface area vector in the outer normal direction of the flow domain. $\mathbf{M}$ is the moment of the pressure $p$ and shear stress $\boldsymbol{\tau}$ over the rotor blade with respect to the rotation axis; however, the contribution from $\boldsymbol{\tau}$ is negligible compared to that from $p$.
The contribution from shear stress on the surfaces excluding the rotor blade surface and the contribution from heat transfer over all surfaces are lumped into $\dot \varepsilon_f$, i.e.,
\begin{equation} \label{eq:shear_stress_work}
    \dot \varepsilon_f = -\int_{\partial \Omega-\mathrm{rotor \ blade}} (\mathbf{V} \cdot \boldsymbol{\tau}) \cdot \mathrm{d}\mathbf{S} + \oint_{\partial \Omega} \mathbf{q} \cdot \mathrm{d}\mathbf{S}
\end{equation}
The contributions of both shear stress and heat transfer vanish on solid surfaces since $\mathbf{V} = 0$ and $\mathbf{q} = 0$ under the no-slip and adiabatic conditions, and their contributions at the turbine inlet and outlet are negligible compared to $P_t$. Consequently, $\dot \varepsilon_f$ is negligible in Eq. (\ref{eq:energy_balance}).
The heat addition rate due to fuel burn in the turbine stage is computed by integrating Eq. (\ref{eq:species}), i.e.,
\begin{equation} \label{eq:overall_heat_release}
    \dot Q_t = \int_\Omega \dot{Q} \mathrm{d}\Omega = -\sum_{i=1}^{N}{h^0_{i} \int_\Omega \dot\omega_i\mathrm{d}\Omega } = -\sum_{i=1}^{N}{ h^0_{i} \oint_{\partial \Omega} \rho Y_i \mathbf{V^\prime} \cdot \mathrm{d}\mathbf{S} } = -\sum_{i=1}^{N}{ h^0_{i} (\dot{m}_{\mathrm{out},i} - \dot{m}_{\mathrm{in},i}) }
\end{equation}

Table \ref{tab:overall_energy_balance_turbine} lists the value of each term in Eq. (\ref{eq:energy_balance}), along with the mass flow rate.  $\dot \varepsilon_f$ is computed by $\dot \varepsilon_f = (\dot{m} \bar H)_{\mathrm{in}} -(\dot{m} \bar H)_{\mathrm{out}} + \dot Q_t - P_t$, which includes the small terms due to viscous forces and heat conduction at the inlet and outlet of the turbine stage in Eq. (\ref{eq:shear_stress_work}) and possible numerical errors in the computation. $\dot \varepsilon_f$ is sufficiently small compared with the turbine power $P_t$, demonstrating that the present computation achieves a good overall energy balance in the turbine stage. The small nonzero values of $\dot Q_t$ in cases 0N and 4N arise from numerical errors introduced during data post-processing.

The small amount of fuel injected at the inlet has little aerodynamic influence in the nonreacting flow. The inlet enthalpy flux is sightly lower in case 4N than in case 0N due to the cold, low-speed fuel injection. However, their outlet enthalpy fluxes are identical. Consequently, the turbine power in case 0N is only 1.3\% higher than that in case 4N. The mass flow rates of the two nonreacting cases are almost identical.

Compared with the nonreacting cases, however, the mass flow rates in cases 4R and 16R decrease by 7\% and 8\%, respectively. This mass-flow reduction is due to the Rayleigh effect of heat addition, which in practice can be compensated by slightly increasing the cross-sectional area of the flow passage. Compared with case 4R, the faster completion of combustion in the stator in case 16R, as shown later, results in more severe thermal choking and thus an additional 1\% reduction in mass flow rate.

Combustion alters the energy balance in the turbine passage. The inlet enthalpy fluxes are almost the same in the two reacting cases, 4R and 16R. Both of them are slightly smaller than those in the two nonreacting cases because of the reduced mass flow rate. 
The overall heat addition is identical for the two reacting cases, indicating that the identical amount of fuel injected at the inlet is completely burned within the turbine passage for both cases. 
The turbine powers in the reacting cases are slightly higher than those in the nonreacting cases, implying the effects of combustion on the aerodynamic loading of turbine blades. Furthermore, the turbine power in case 16R is further increased compared with that in case 4R.  
Although the turbine power outputs are close between the nonreacting and reacting cases, the reduced mass flow rates in the reacting cases result in higher turbine powers per unit mass flow rate, as will be analyzed in \ref{sec:thermodynamic_performance}.

\begin{table}[htb!]
    \centering
    \caption{Overall energy balance in a turbine-stage passage (Units: kW)}
    \begin{tabular}{ccccccc}
        \hline
          Case & $(\dot{m} \bar H)_{\mathrm{in}}$  & $(\dot{m} \bar H)_{\mathrm{out}}$  & $\dot Q_t$ & $P_t$ & $\dot \varepsilon_f$ & $\dot{m}$, kg/s \\
        \hline
        0N  &  2889   &  2259   &  -1.3   &  626.6   &  2.1   & 1.745 \\
        4N  &  2875   &  2257   &  -8.4   &  618.4   &  -8.8  & 1.747 \\
        4R  &  2680   &  2645   &  591.2  &  633.4   &  -7.2  & 1.626 \\
        16R &  2667   &  2609   &  593.9  &  646.6   &  5.3   & 1.615 \\
        \hline
    \end{tabular}
    \label{tab:overall_energy_balance_turbine}
\end{table}

\begin{figure}[htb!]
    \centering
    \begin{subfigure}[b]{0.49\linewidth}
        \centering
        \includegraphics[width=1\linewidth]{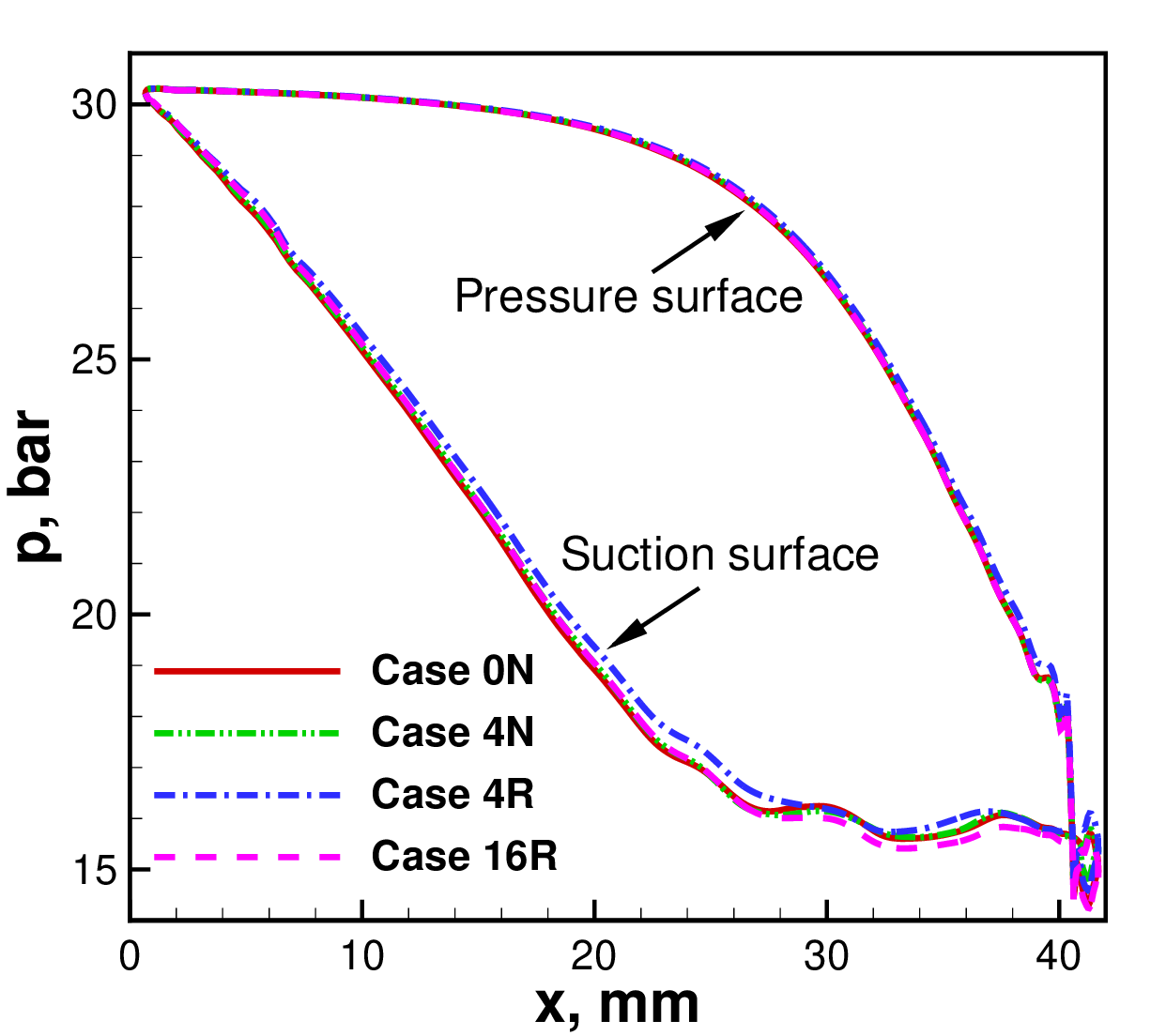}
        \caption{Stator}
        \label{fig:p-x_blade_stator}
    \end{subfigure}
    \begin{subfigure}[b]{0.49\linewidth}
        \centering
        \includegraphics[width=1\linewidth]{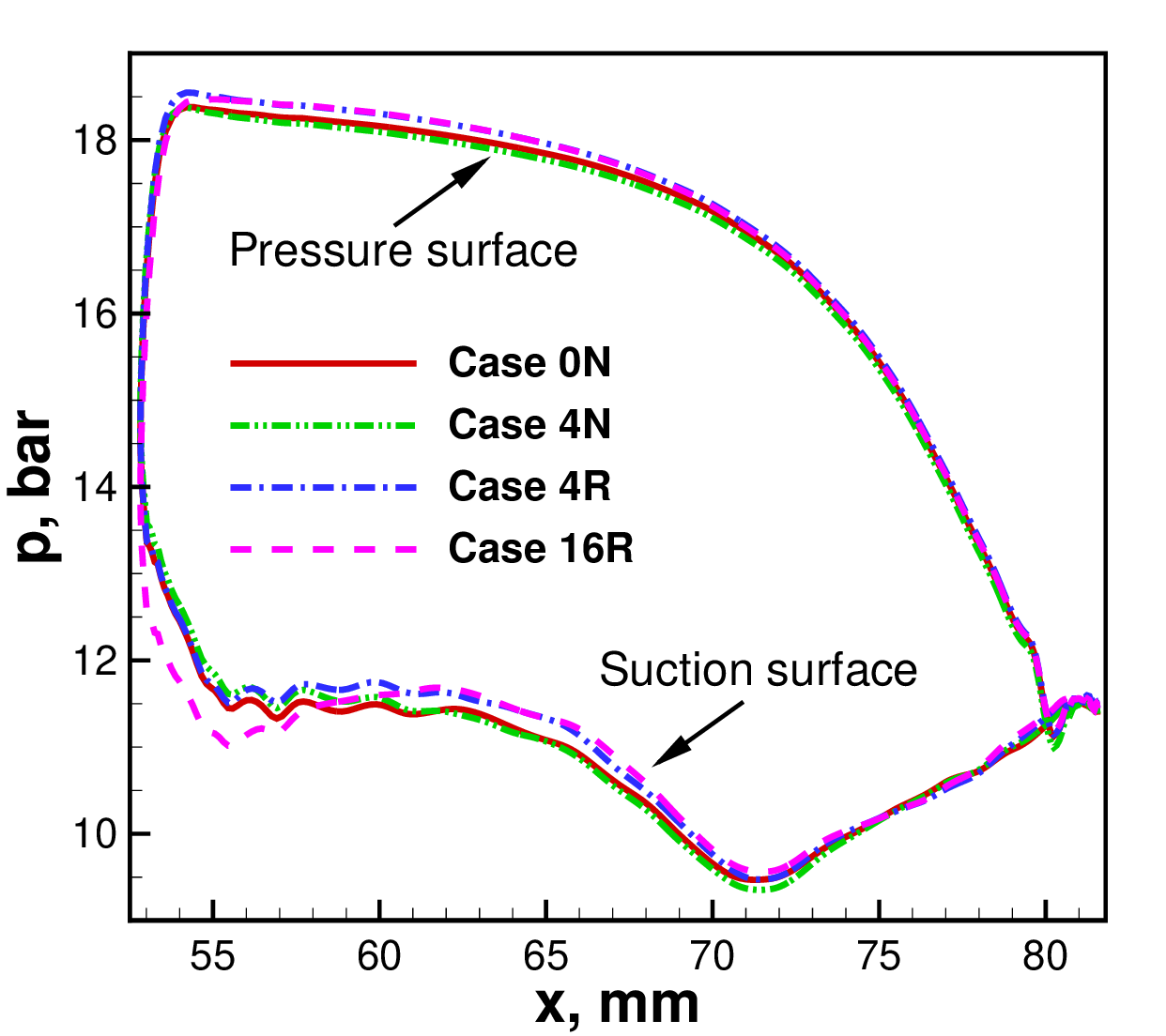}
        \caption{Rotor}
        \label{fig:p-x_blade_rotor}
    \end{subfigure}
    \caption{Axial variation of spanwise-averaged static pressure over the blade. The stator leading edge is at x = 0.}
    \label{fig:p-x_blade_nonreacting_reacting}
\end{figure}

The power increases in the reacting cases can be revealed by examining the pressure distributions over the blade surfaces, as indicated by Eq. (\ref{eq:work_power}). The variations of static pressure along the stator and rotor blade chords are shown in Fig. \ref{fig:p-x_blade_nonreacting_reacting}, where the time-averaged pressure is further averaged along the blade span. 
Close pressure distributions over the stator blade in all the cases indicate that fuel injection and chemical reactions have little influence on the aerodynamic loading of the stator blade. This is because the fuel jets and flames go through the stator passage far away from the blade surface, as shown in Fig. \ref{fig:contour_TA_radial_slice}. 
The pressure over the suction surface in case 4R is slightly higher than that in case 16R. This behavior may be attributed to the thicker flames generated by the larger fuel injectors in case 4R, which pass closer to the blade within the stator passage and consequently produce the higher temperature and thus higher pressure over the stator blade surface.

\begin{figure}[htb!]
    \centering
    \begin{subfigure}[b]{0.49\linewidth}
        \centering
        \includegraphics[width=1\linewidth]{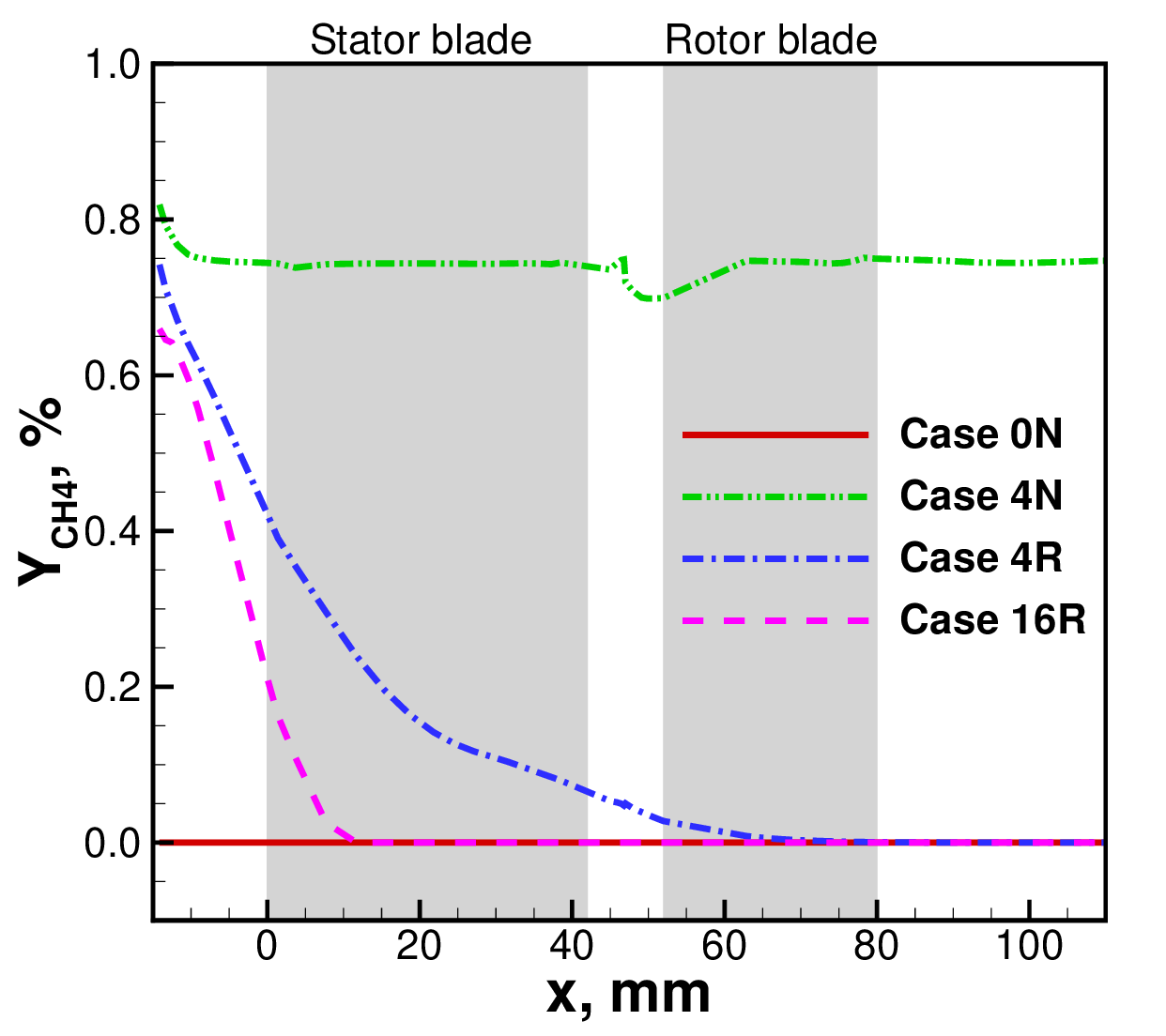}
        \caption{Mass fraction of CH\textsubscript{4}}
        \label{fig:CH4-x_nonreacting_reacting}
    \end{subfigure}
    \begin{subfigure}[b]{0.49\linewidth}
        \centering
        \includegraphics[width=1\linewidth]{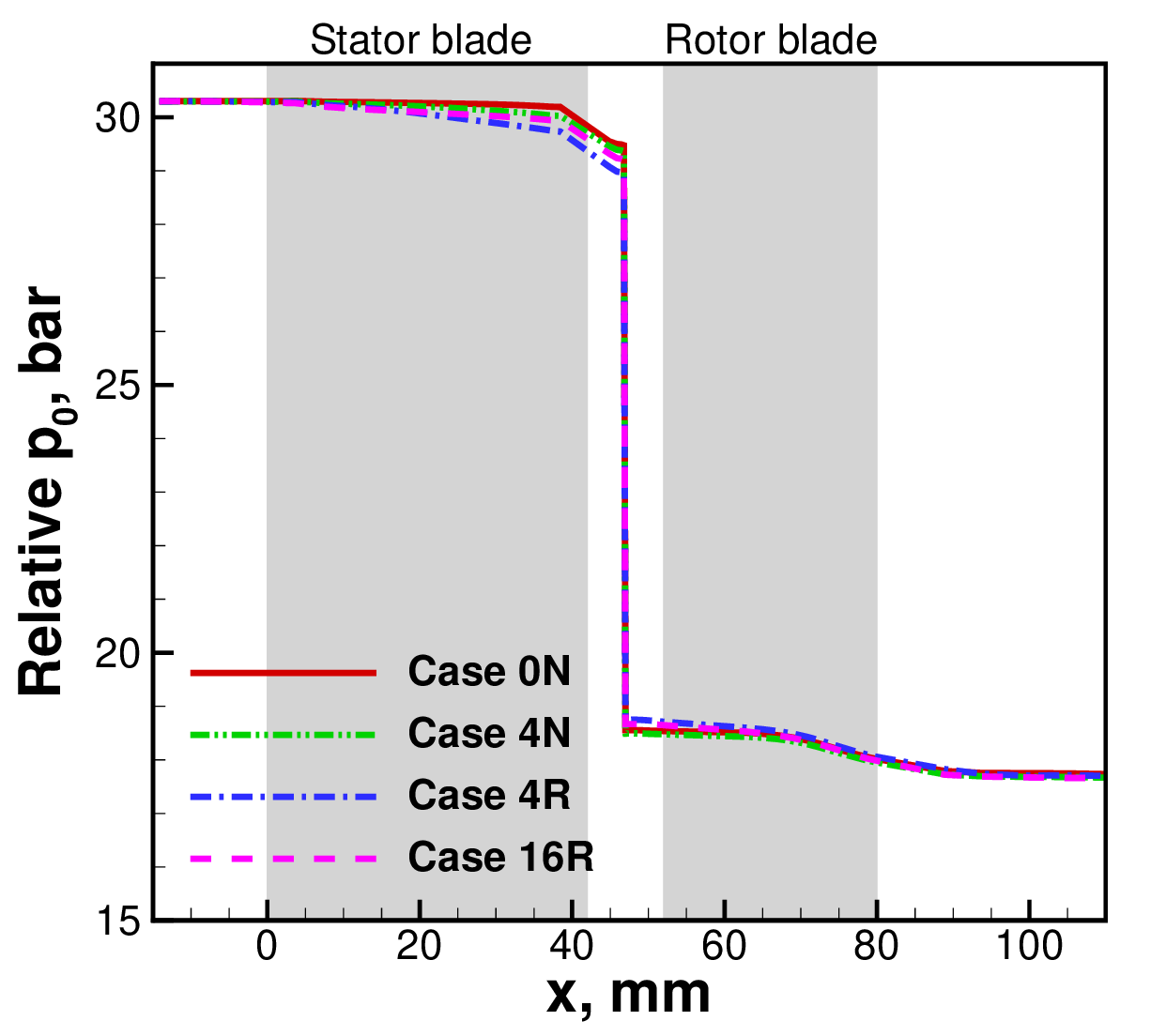}
        \caption{Relative total pressure}
        \label{fig:P0-x_nonreacting_reacting}
    \end{subfigure}
    \begin{subfigure}[b]{0.49\linewidth}
        \centering
        \includegraphics[width=1\linewidth]{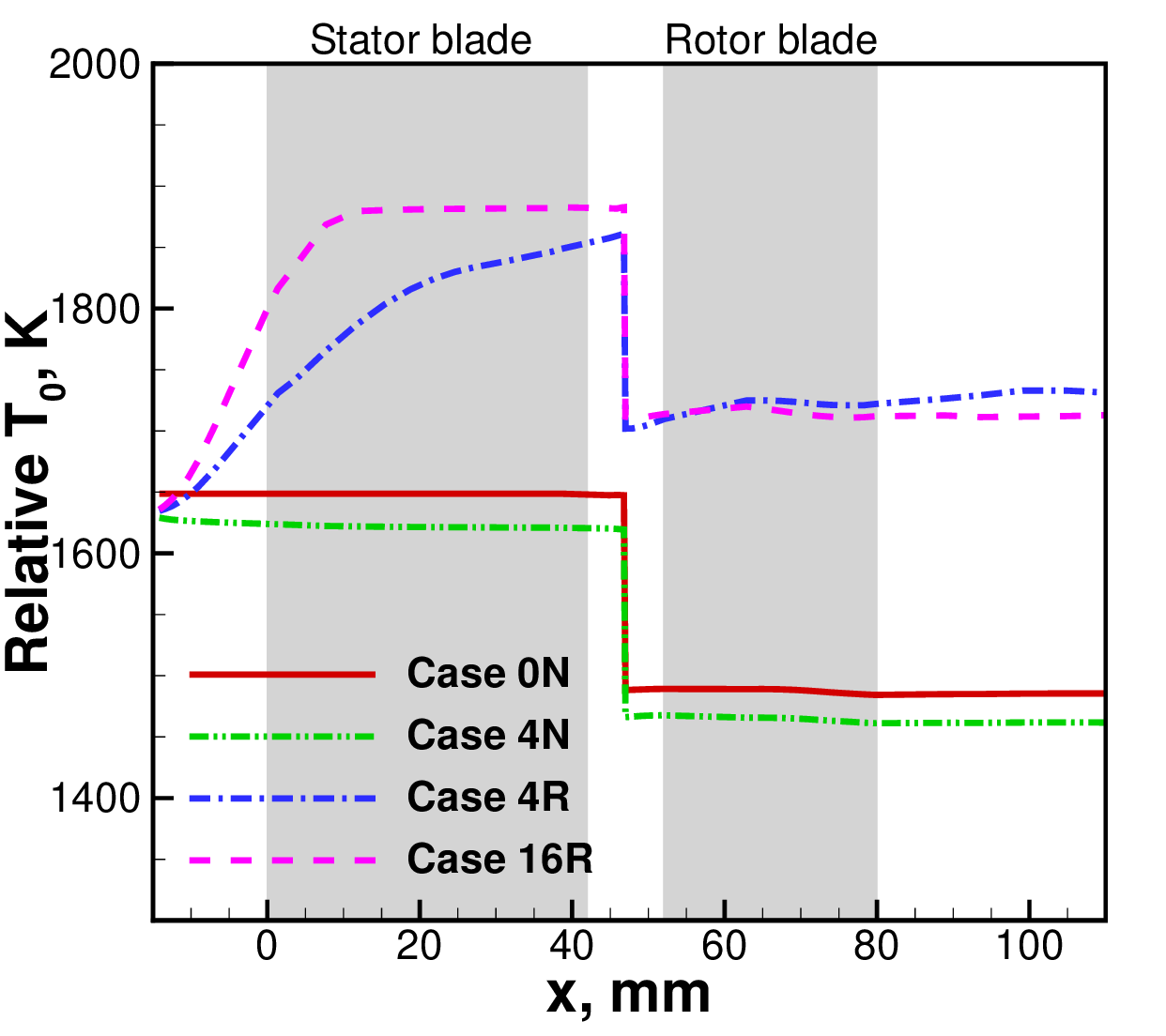}
        \caption{Relative total temperature}
        \label{fig:T0-x_nonreacting_reacting}
    \end{subfigure}
    \begin{subfigure}[b]{0.49\linewidth}
        \centering
        \includegraphics[width=1\linewidth]{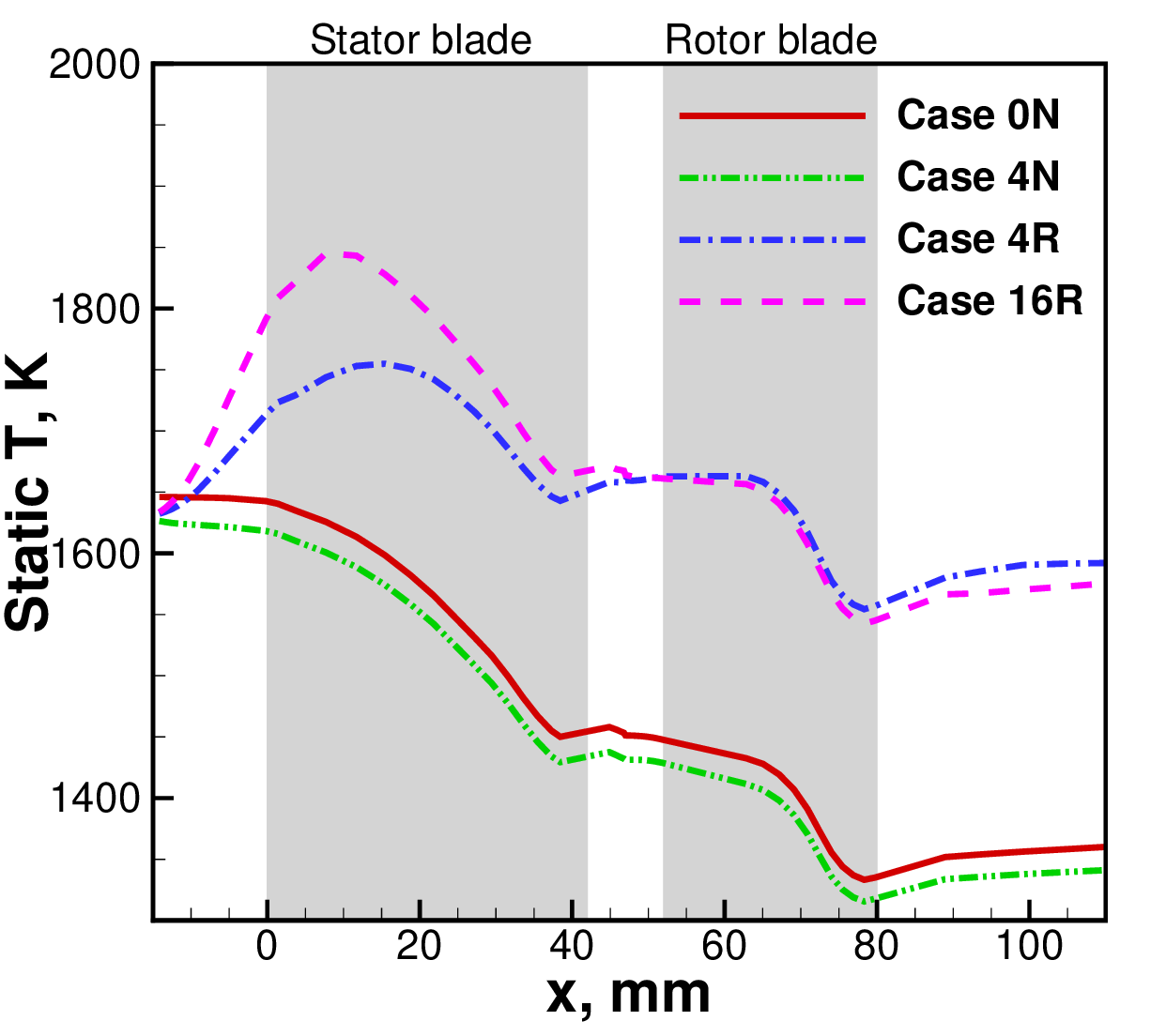}
        \caption{Static temperature}
        \label{fig:T-x_nonreacting_reacting}
    \end{subfigure}
    \caption{Axial variations of surface-averaged flow quantities. The stator leading edge is at x = 0.}
    \label{fig:P0_T0_CH4-x_nonreacting_reacting}
\end{figure}

However, different behaviors appear in the rotor. A comparison between cases 4N and 4R indicates that combustion raises the static pressure on the first two-thirds of the pressure surface and the middle half of the suction surface. This occurs because the flames and hot streaks from the stator first impinge on the rotor pressure surface and then are convected towards the suction surface of the adjacent blade, as shown in Fig. \ref{fig:contour_instant_radial_slice_T}.
The larger pressure difference between the suction and pressure surfaces in case 4R produces a higher aerodynamic loading of the rotor and thus an increased turbine power than in the corresponding nonreacting case. 
The pressure distributions in cases 4R and 16R are almost the same on the rotor blade, except on the first third of the suction surface, where the static pressure in case 16R is obviously lower. This difference is attributed to the distinct burning behaviors in the two cases. In case 4R, strong reactions occur near the leading edge of the rotor suction surface, as shown in Fig. \ref{fig:contour_instant_radial_slice_FI}. In contrast, all injected fuel is consumed within the upstream stator passage in case 16R, as will be shown in Fig. \ref{fig:P0_T0_CH4-x_nonreacting_reacting}. This results in a lower temperature level in case 16R and thus a reduced pressure near the front region of the rotor suction surface.
This pressure reduction in case 16R contributes to its higher turbine power, as listed in Table \ref{tab:overall_energy_balance_turbine}.

Figure \ref{fig:P0_T0_CH4-x_nonreacting_reacting} shows the axial variations of surface-averaged CH\textsubscript{4} mass fraction, relative total pressure, relative total temperature, and static temperature. The axial regions of the stator and rotor blades are shaded in gray.
The CH\textsubscript{4} mass fraction is exactly zero in case 0N because no fuel is injected at the inlet. In case 4N, it almost remains a constant, 0.75\%, throughout the turbine, as expected. The slight fluctuations near the stator and rotor inlets are attributed to data-processing errors associated with strong mixing of fuel and air within finite grid resolution.
In cases 4R and 16R, the injected fuel is consumed by chemical reactions, as indicated by the reducing CH\textsubscript{4} mass fraction along the axial direction. In case 4R, the fuel is fully depleted around the mid-chord of the rotor blade. However, in case 16R, the fuel is consumed by the first third of the stator blade, producing a steeper slope. 

A relative quantity is defined in the reference frame of the rotor, in contrast to an absolute quantity which is defined in the stationary reference frame. Naturally, the absolute and relative quantities in a stator are equivalent. A relative quantity exhibits an abrupt change across the stator-rotor interface as the contribution of rotational velocity is removed when the quantity is defined in the relative reference frame.
In a conventional turbine without burning, the relative total temperature is theoretically conserved across both stator and rotor, while the relative total pressure gradually decreases due to losses.
The difference in relative total temperature between the reacting and nonreacting cases reflects the energy addition from chemical reactions, while the difference in relative total pressure represents the loss induced by those reactions.

In the present turbine stage, the relative total pressure in all cases gradually decreases across both the stator and rotor passages, as expected. In case 0N, the baseline case, it drops about 2.7\% across the stator and 4.6\% across the rotor.
At a specific axial position, the relative total pressures among the cases are extremely close, with the maximum difference less than 1.5\%. This implies that the fuel injection and chemical reactions within the turbine have minimal influence on the overall loss in it.

In the two nonreacting cases, 0N and 4N, the relative total temperature remains constant across both the stator and rotor passages, as expected. The total temperature in case 4N is slightly lower because of the cold, low-speed fuel injected at the inlet. 
In case 4R, the relative total temperature increases throughout the stator and the first half of the rotor, with a gradually decreasing slope, and remains approximately constant in the downstream rotor. These variations are consistent with the behaviors of strong reactions near the stator inlet, progressively weakened reactions downstream, and the quenched state in the downstream rotor, as shown in Figs. \ref{fig:contour_instant_radial_slice}, \ref{fig:contour_instant_stream_slice}, and \ref{fig:contour_TA_radial_slice}.
In contrast, the relative total temperature in case 16R rises rapidly before the first third of the stator and then remains nearly constant in the downstream stator and throughout the rotor. Consequently, this case maintains a higher total temperature level across the entire stator passage than case 4R. These observations indicate earlier and faster chemical reactions in case 16R than in case 4R. 
This is also evidenced by the variations of CH\textsubscript{4} mass fractions in Fig. \ref{fig:CH4-x_nonreacting_reacting}.
The higher total temperature at the stator outlet in case 16R compared with case 4R indicates a greater potential to extract work in the downstream rotor, given that both cases experience a nearly identical total pressure drop across the rotor. This implies that the turbine specific work (power per unit mass flow rate) in case 16R exceeds that in case 4R, as will be shown in Sec. \ref{sec:thermodynamic_performance}.
This provides an instruction to turbine-burner designers that early completion of combustion in the upstream stator can enhance the work extraction in the downstream rotor. 

As shown in Fig. \ref{fig:T-x_nonreacting_reacting}, the static temperature decreases across both the stator and rotor passages in the two nonreacting cases, as thermal energy is converted into kinetic energy within the stator and mechanical work is extracted by the rotor blades.
Although the temperature levels in the two reacting cases are approximately $200 \, \mathrm{K}$ higher than those in the nonreacting cases, they still decrease along the axial direction, except in the upstream region of the stator passage where intense combustion occurs.
In addition, for the reacting cases, the temperature within the rotor passage remains below the turbine inlet temperature, which is beneficial for alleviating the thermal loading on the downstream rotor in a turbine-burner configuration.

\begin{figure}[htb!]
    \centering
    \begin{subfigure}[b]{0.25\linewidth}
        \centering
        \includegraphics[width=1\linewidth]{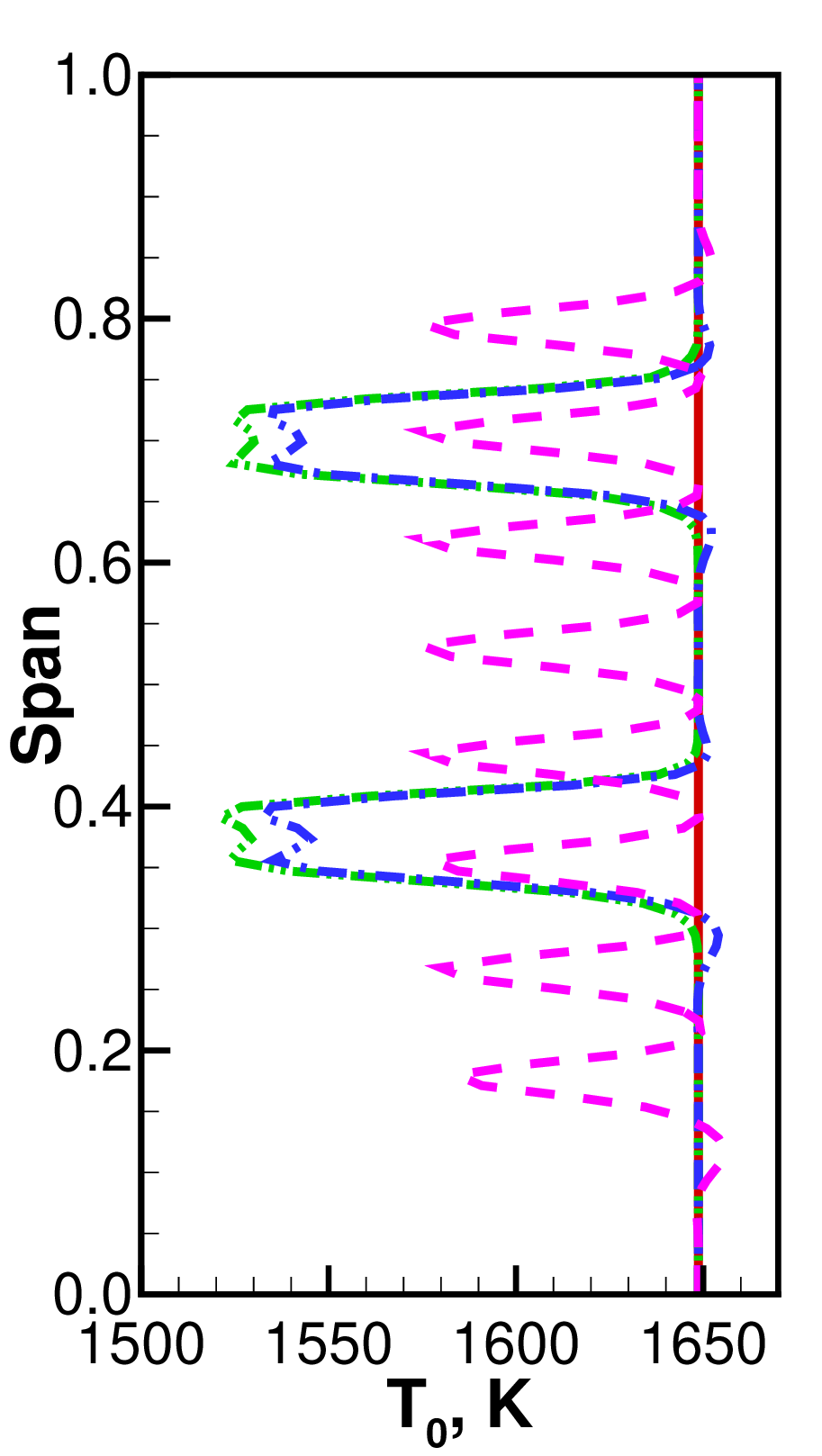}
        \caption{Total temperature}
        \label{fig:spandistr_T0_stator_inlet}
    \end{subfigure}
    \begin{subfigure}[b]{0.25\linewidth}
        \centering
        \includegraphics[width=1\linewidth]{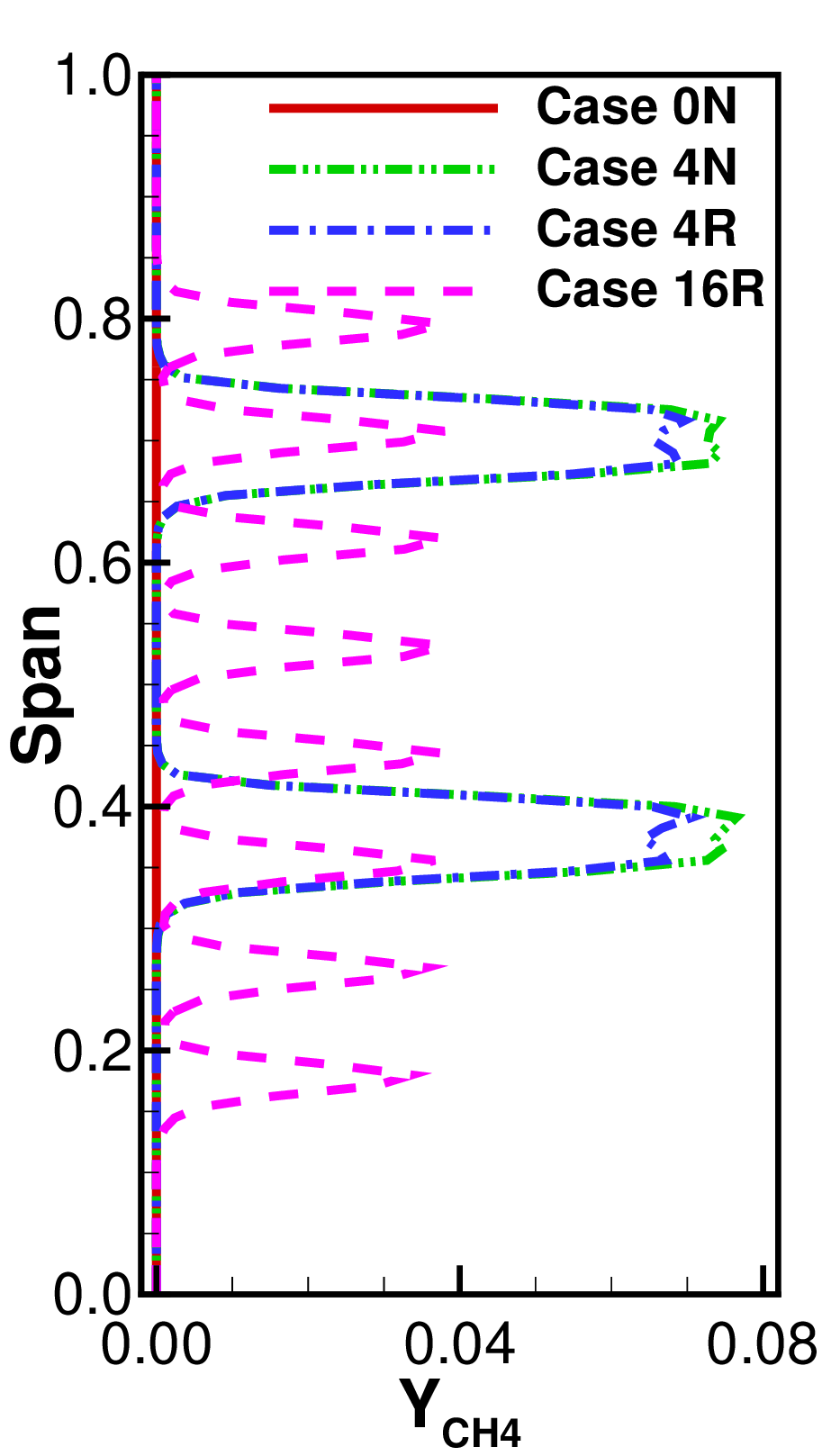}
        \caption{Mass fraction of CH\textsubscript{4}}
        \label{fig:spandistr_CH4_stator_inlet}
    \end{subfigure}
    \begin{subfigure}[b]{0.25\linewidth}
        \centering
        \includegraphics[width=1\linewidth]{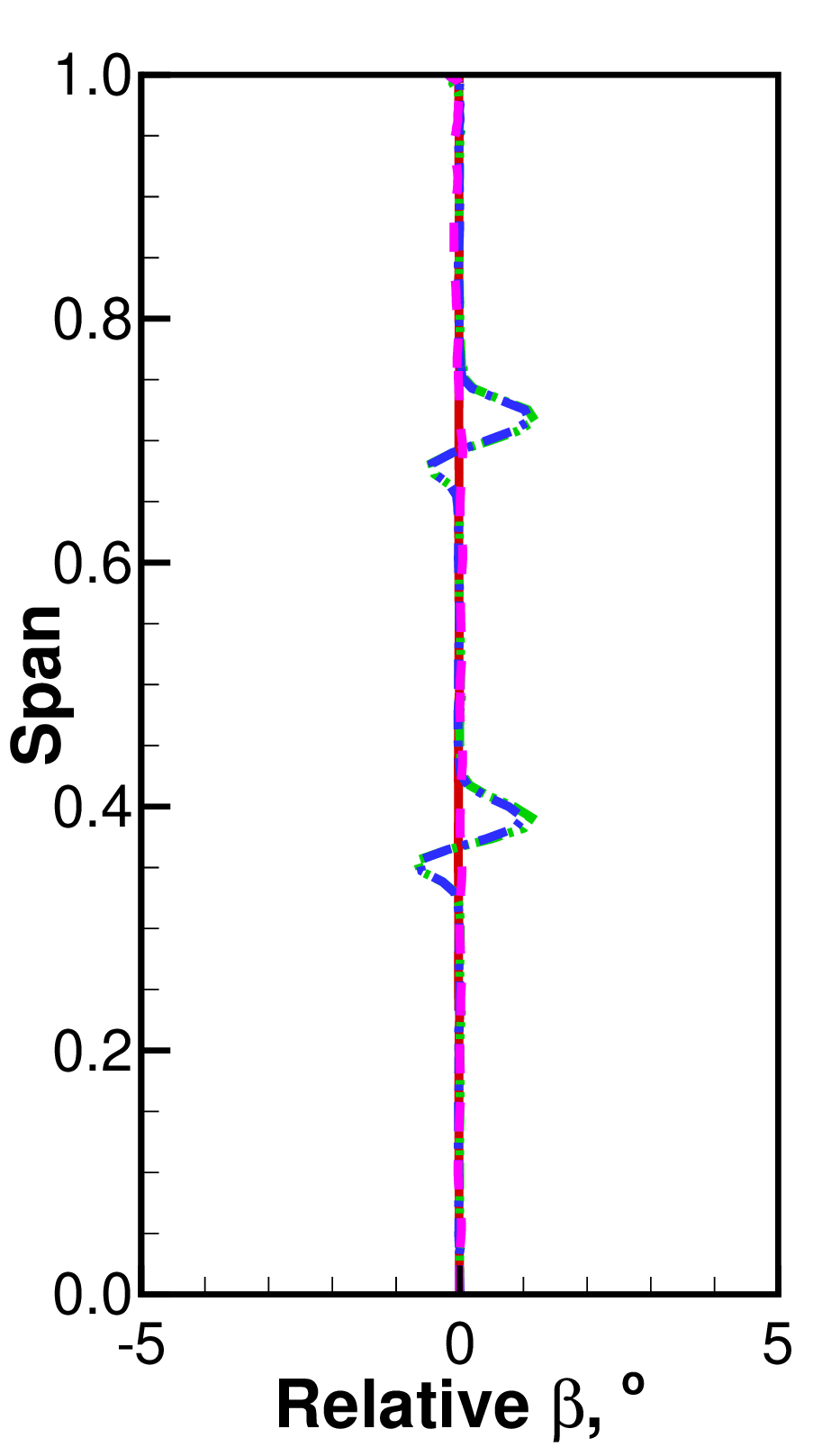}
        \caption{Relative flow angle}
        \label{fig:spandistr_beta_stator_inlet}
    \end{subfigure}
    \caption{Spanwise variations of flow parameters at stator inlet.}
    \label{fig:spandistr_nonreacting_reacting_stator_inlet}
\end{figure}

\begin{figure}[htb!]
    \centering
    \begin{subfigure}[b]{0.25\linewidth}
        \centering
        \includegraphics[width=1\linewidth]{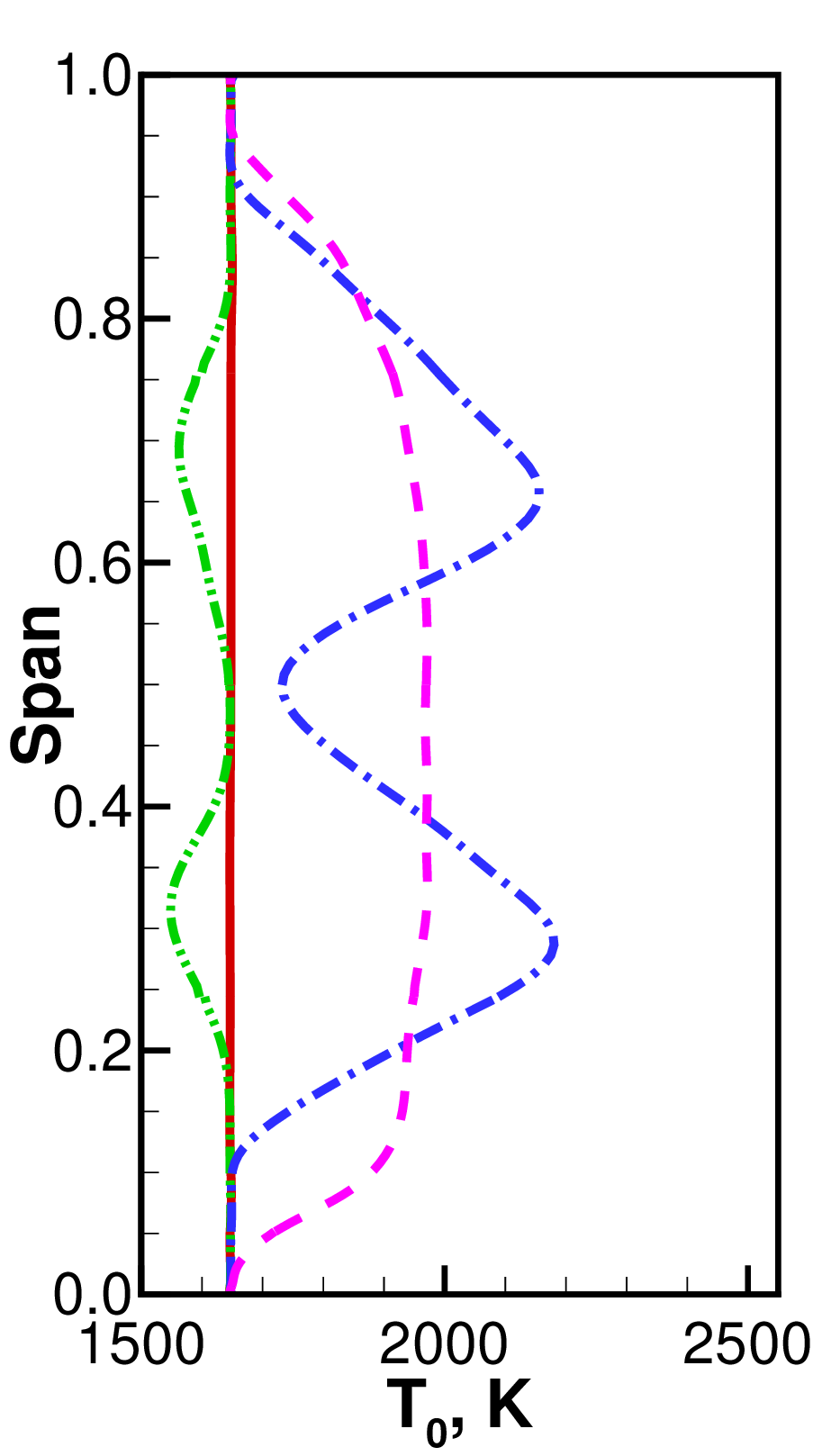}
        \caption{Total temperature}
        \label{fig:spandistr_T0_stator_outlet}
    \end{subfigure}
    \begin{subfigure}[b]{0.25\linewidth}
        \centering
        \includegraphics[width=1\linewidth]{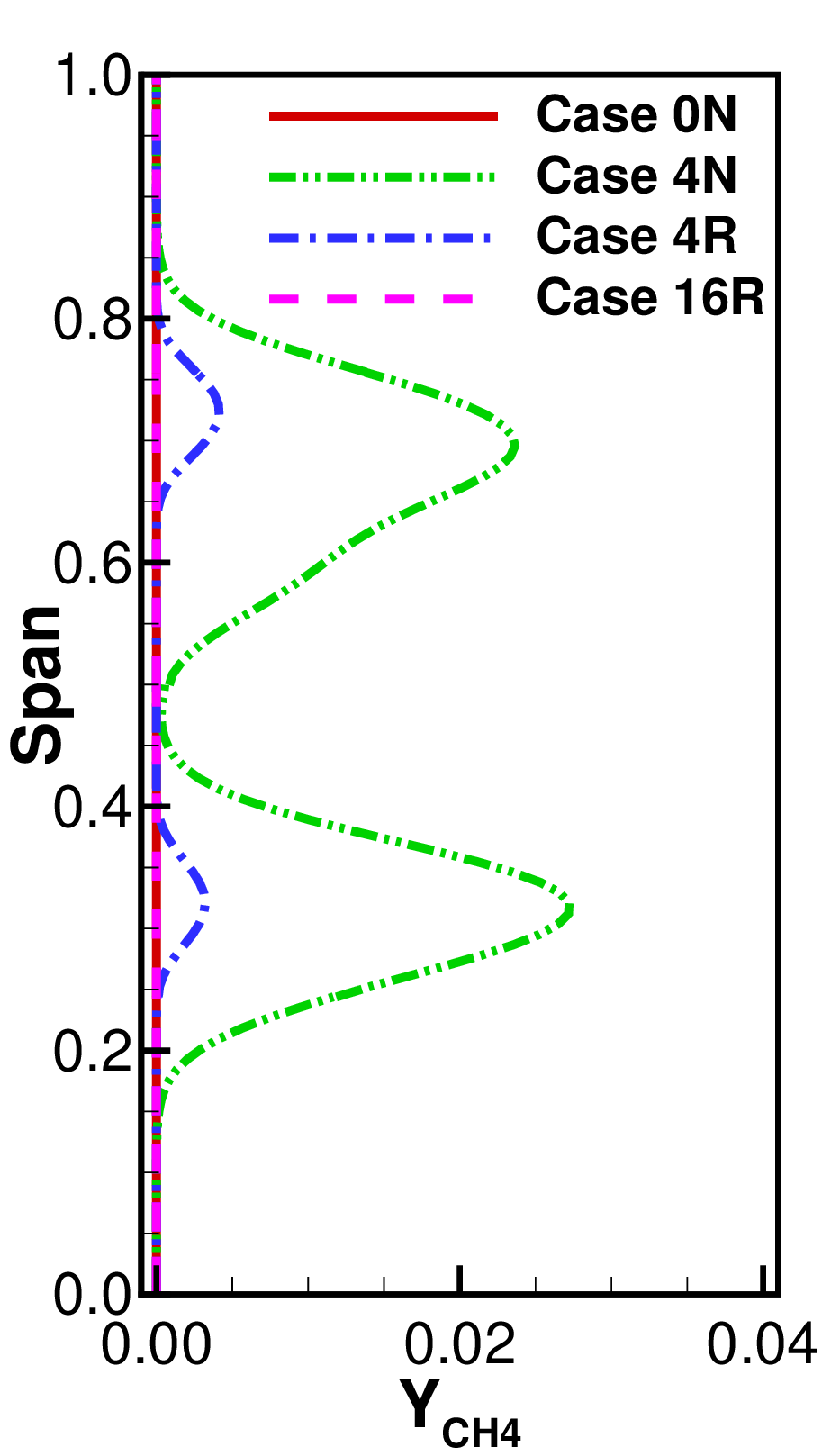}
        \caption{Mass fraction of CH\textsubscript{4}}
        \label{fig:spandistr_CH4_stator_outlet}
    \end{subfigure}
    \begin{subfigure}[b]{0.25\linewidth}
        \centering
        \includegraphics[width=1\linewidth]{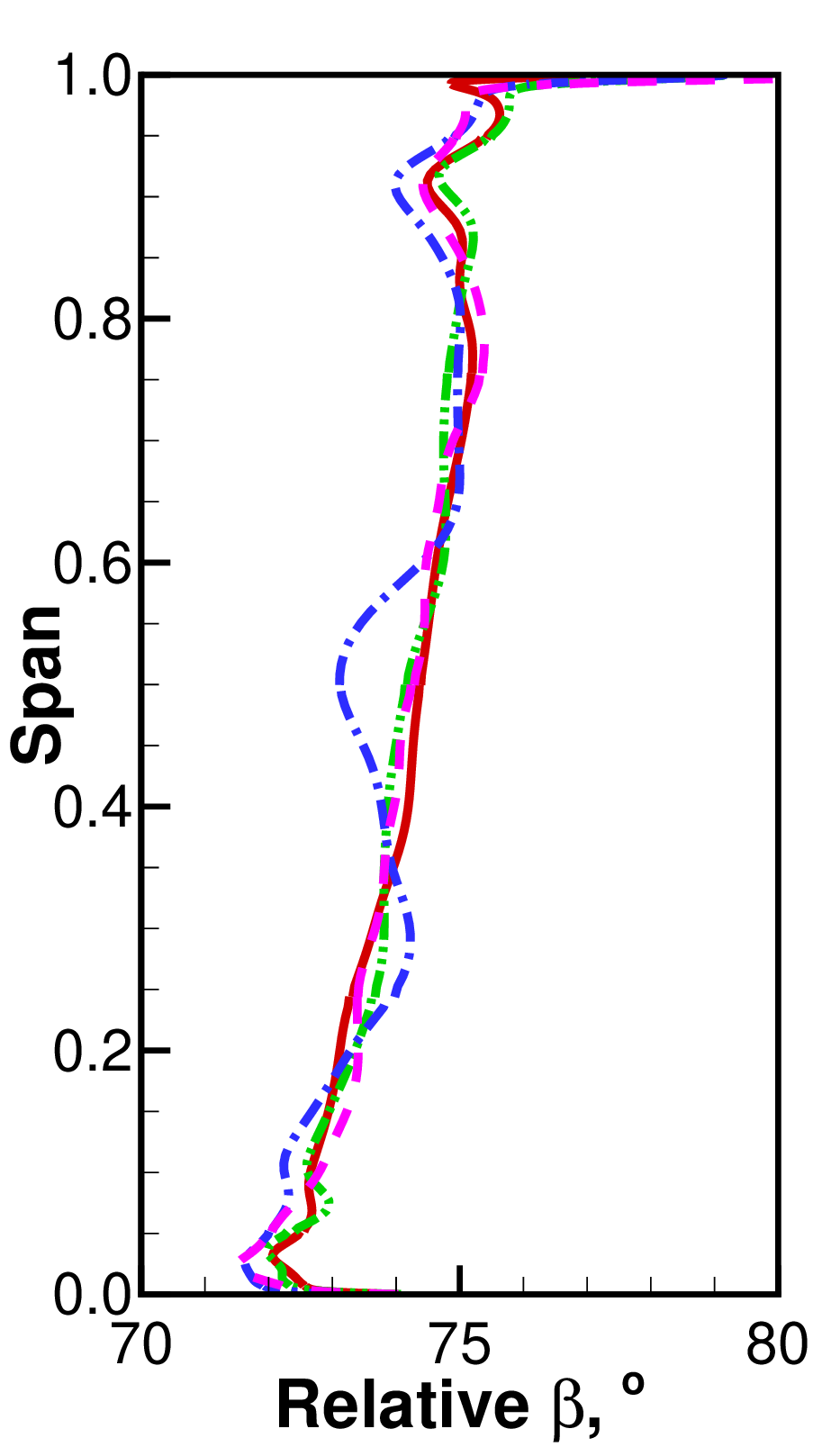}
        \caption{Relative flow angle}
        \label{fig:spandistr_beta_stator_outlet}
    \end{subfigure}
    \caption{Spanwise variations of flow parameters at stator outlet.}
    \label{fig:spandistr_nonreacting_reacting_stator_outlet}
\end{figure}

\begin{figure}[htb!]
    \centering
    \begin{subfigure}[b]{0.25\linewidth}
        \centering
        \includegraphics[width=1\linewidth]{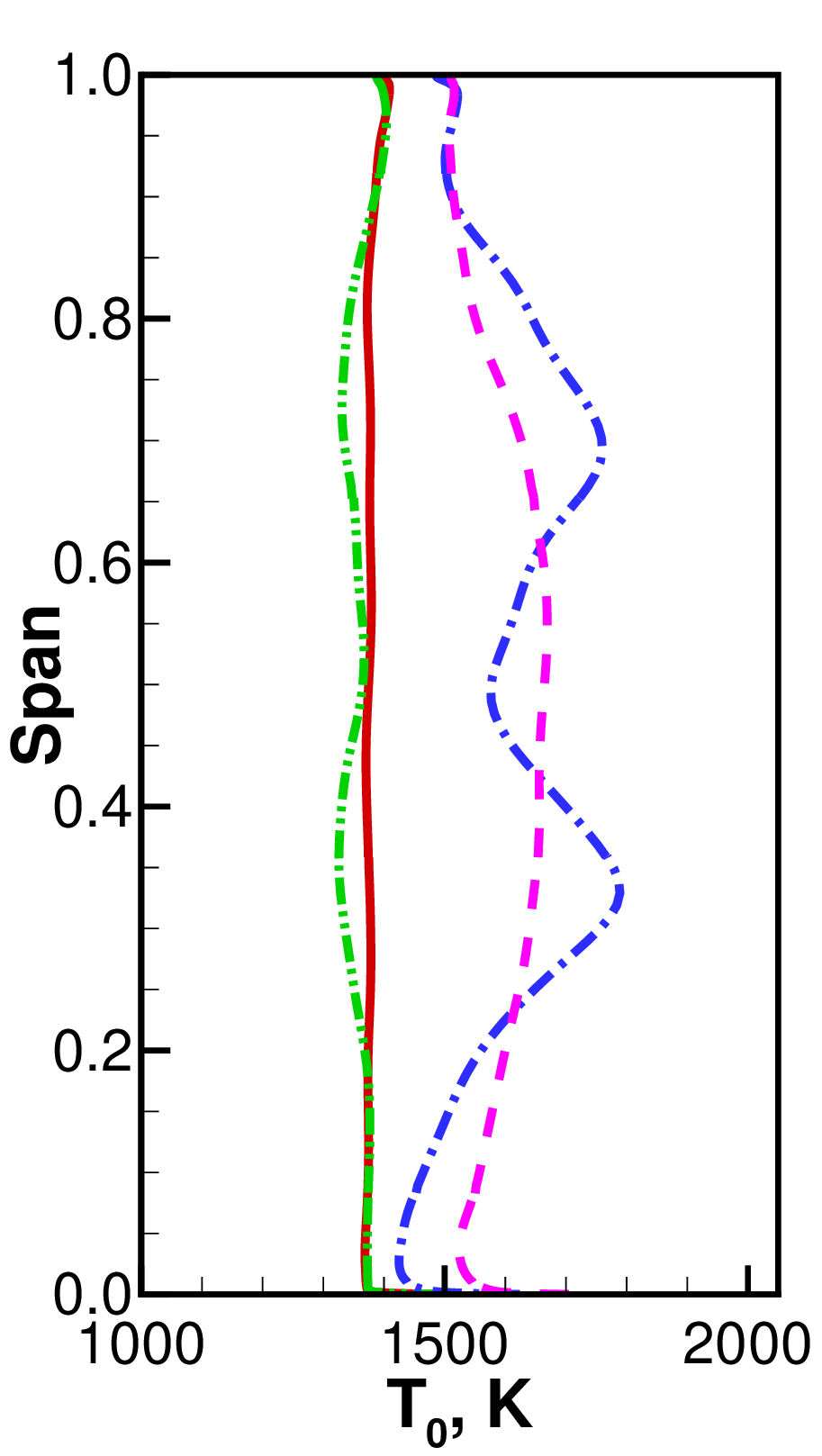}
        \caption{Total temperature}
        \label{fig:spandistr_T0_rotor_outlet}
    \end{subfigure}
    \begin{subfigure}[b]{0.25\linewidth}
        \centering
        \includegraphics[width=1\linewidth]{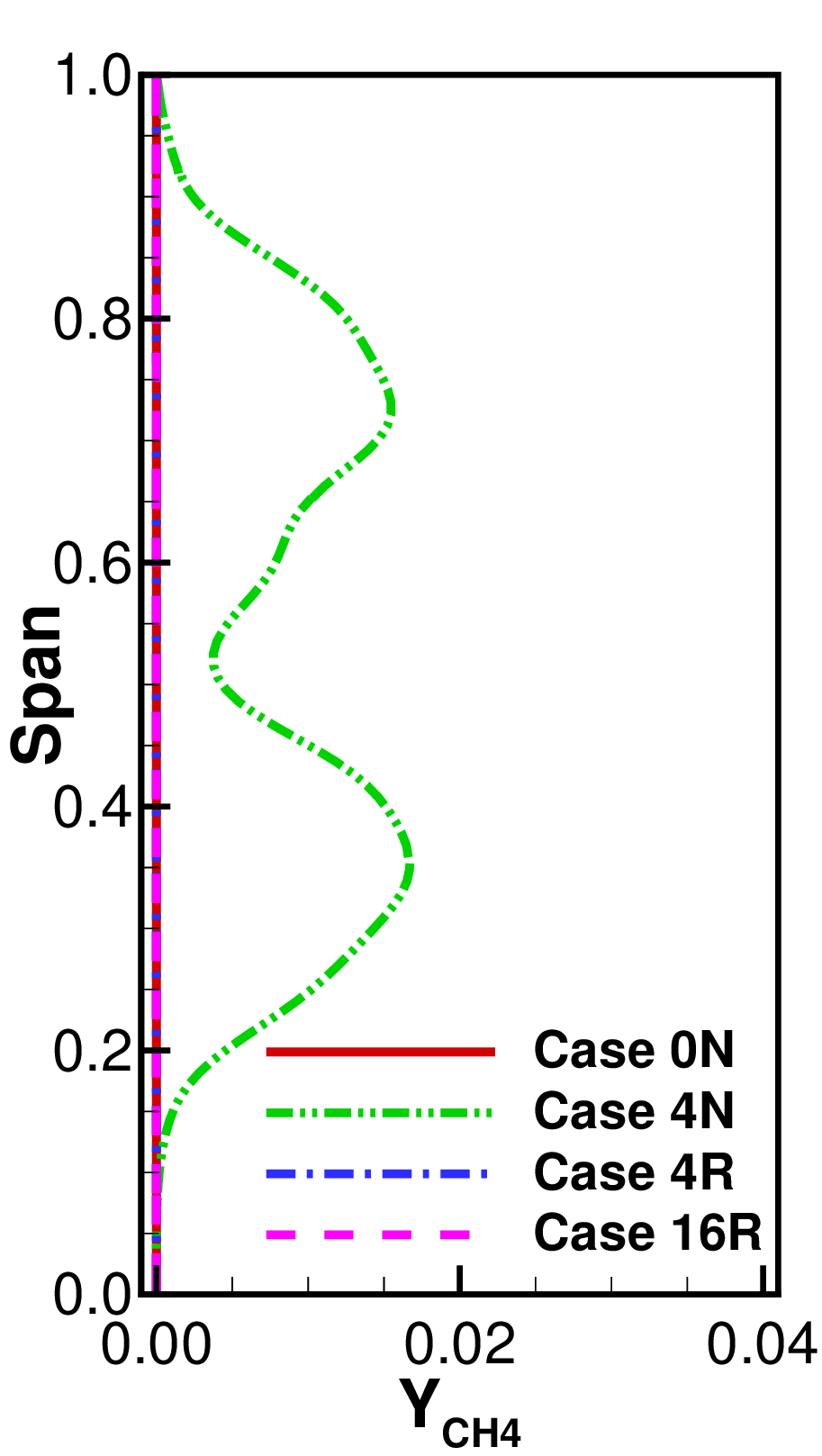}
        \caption{Mass fraction of CH\textsubscript{4}}
        \label{fig:spandistr_CH4_rotor_outlet}
    \end{subfigure}
    \begin{subfigure}[b]{0.25\linewidth}
        \centering
        \includegraphics[width=1\linewidth]{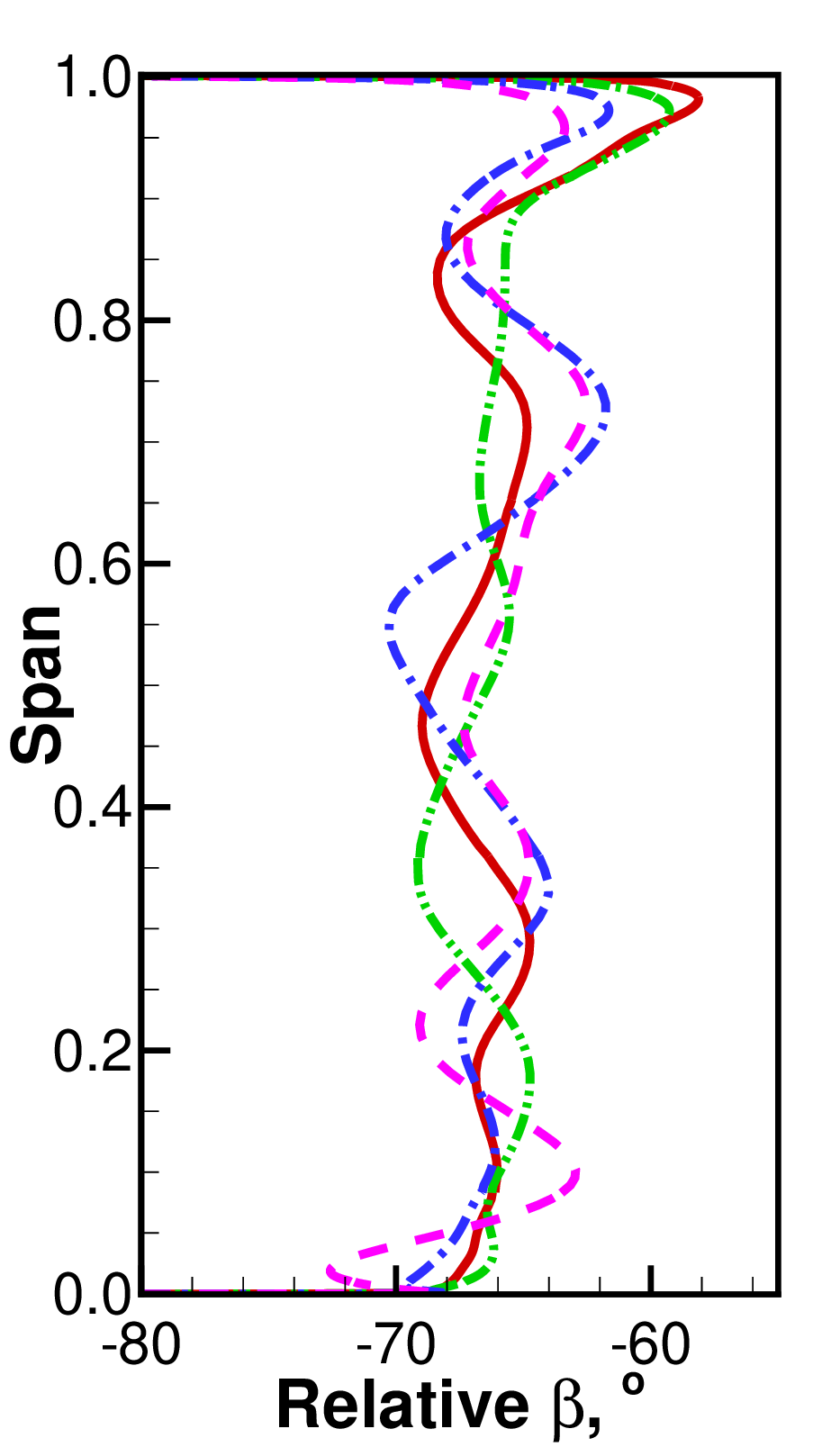}
        \caption{Relative flow angle}
        \label{fig:spandistr_beta_rotor_outlet}
    \end{subfigure}
    \caption{Spanwise variations of flow parameters at rotor outlet.}
    \label{fig:spandistr_nonreacting_reacting_rotor_outlet}
\end{figure}

Figures \ref{fig:spandistr_nonreacting_reacting_stator_inlet}, \ref{fig:spandistr_nonreacting_reacting_stator_outlet}, and \ref{fig:spandistr_nonreacting_reacting_rotor_outlet} show the spanwise variations of flow parameters at the turbine inlet, the stator outlet, and the rotor outlet, respectively. 
The spanwise position is normalized as $\bar r = (r - r_h)/(r_c - r_h)$, where $r_h$ and $r_c$ are the hub and casing radii, respectively. At each spanwise position, the flow parameters are obtained by mass-weighted averaging of the time-averaged variables over one pitch.
The relative flow angle $\beta$ is defined as the angle between the relative velocity vector and the axial direction and is equivalent to the absolute flow angle in the stator.
At the turbine inlet, the total-temperature profiles in cases 4N and 4R exhibit two deficits caused by the cold, low-speed fuel jets, whereas case 16R shows eight such deficits. Correspondingly, the CH\textsubscript{4} mass-fraction profiles contain two and eight peaks, respectively. The mass-fraction peaks in case 4R are slightly lower than those in case 4N, consistent with the overall mass-fraction values at the inlet shown in Fig. \ref{fig:CH4-x_nonreacting_reacting}. Consequently, the total temperature deficits in case 4R are slightly higher than those in case 4N. Disturbed by the pre-swirling of the fuel jets at the inlet, the flow angle in cases 4N and 4R shows one-degree deviation in the injector regions. 

Apparent differences between the nonreacting and reacting cases appear in the downstream. 
The spanwise diffusion of the fuel jets in case 4R is faster than that in case 4N, as shown by the total-temperature profiles in Figs. \ref{fig:spandistr_T0_stator_outlet} and \ref{fig:spandistr_T0_rotor_outlet}. This trend is consistent with the results in a mixing layer \cite{zhu2024numerical}. As a result, the hub-injector flames in case 4R almost touch with the casing-injector flames at the stator outlet, as indicated by the distribution of total temperature. Actually, they completely meet with each other at the mid-chord of the rotor blade, as shown in Fig. \ref{fig:contour_instant_stream_slice}.
Furthermore, the spanwise diffusion of fuel jets and thus chemical reactions is faster in case 16R than in case 4R, indicated by the zero CH\textsubscript{4} mass fraction and nearly uniform total-temperature profile at the stator outlet. Compared with case 4R, although case 16R shows a higher surface-averaged total temperature at the stator outlet (see Fig. \ref{fig:T0-x_nonreacting_reacting}), the peak value in the total-temperature profile is approximately $200 \, \mathrm{K}$ lower. Similarly, the peak total temperature at the rotor outlet in case 16R is $100 \, \mathrm{K}$ lower. Such a uniform spanwise distribution of total temperature is beneficial to the aerodynamic performance in a well-designed turbine, such as improving the rotor work output.

In contrast to the total temperature and CH\textsubscript{4} mass fraction, the flow angle at the stator outlet in Fig. \ref{fig:spandistr_beta_stator_outlet} is almost the same among all the cases, except that case 4R shows slight fluctuations in the mid-span region. 
At the rotor outlet, slight differences among the cases are observed at specific spanwise locations, as shown in Fig. \ref{fig:spandistr_beta_rotor_outlet}. However, the mean flow angles along the spanwise direction are nearly the same for all cases.
These behaviors demonstrate that fuel injection and chemical reaction have little impact on the flow angle. This is because the outlet flow angle in a turbine stator or rotor is dominantly determined by its metal angle within a wide range of operating conditions.

\begin{figure}[htb!]
    \centering
    \includegraphics[width=0.49\linewidth]{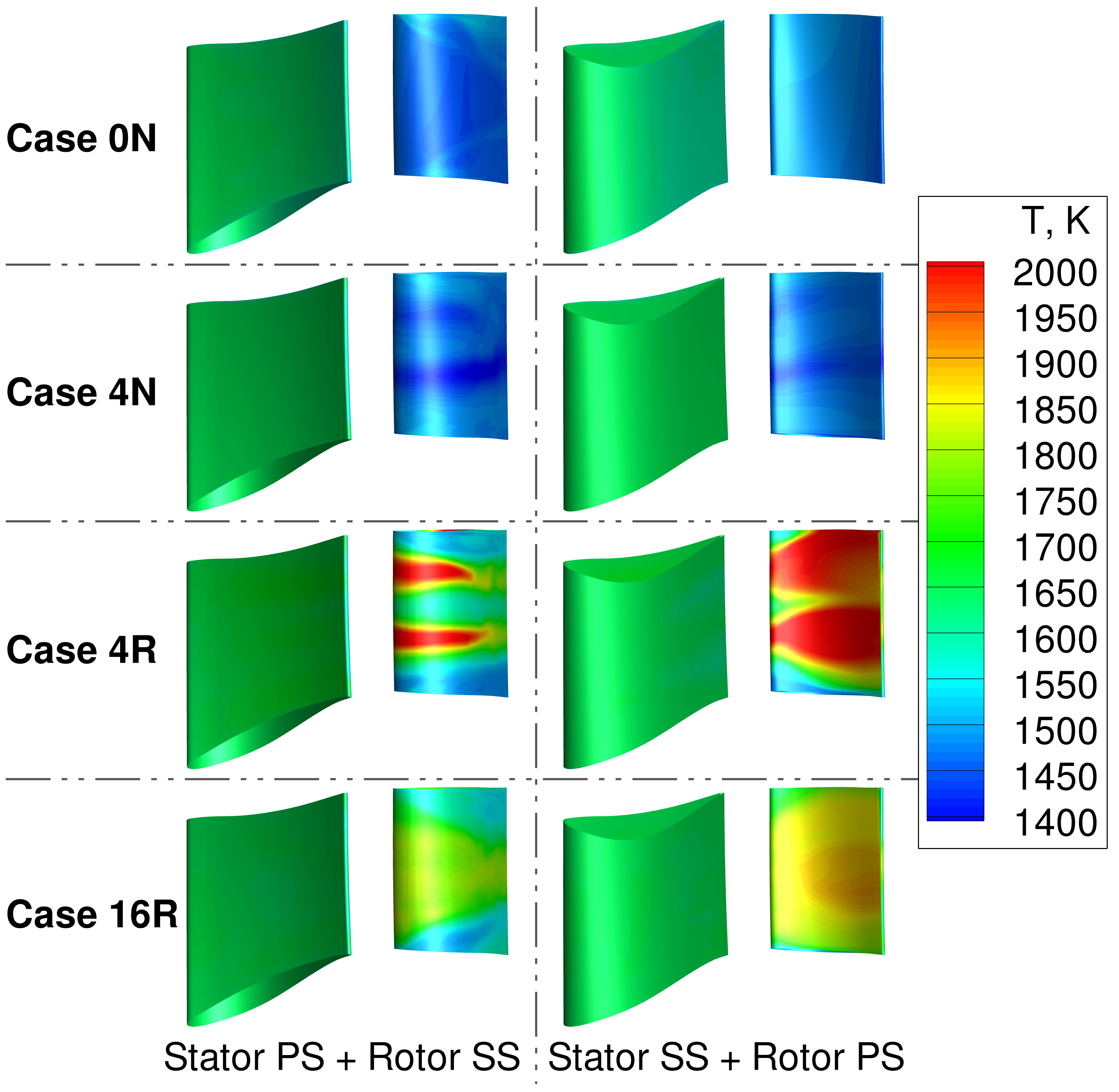}
    \caption{Contours of time-averaged temperature over blade surfaces. SS: suction surface; PS: pressure surface.}
    \label{fig:contour_T_blade_surface}
\end{figure}

Fuel injection and chemical reaction within the turbine affect the temperature over the blades. In the present study, an adiabatic wall boundary condition is assumed. Figure \ref{fig:contour_T_blade_surface} shows the contours of time-averaged temperature over the stator and rotor blade surfaces. 
The temperature levels on the stator blade, including both the suction and pressure surfaces, are almost the same among all the cases, indicating that the fuel injection and chemical reaction have little effects on the temperature over the stator blade surfaces. This is expected since the fuel injectors at the turbine inlet are deliberately arranged to pass through the stator passage away from both the suction and pressure surfaces. 

However, flames and hot streaks originating in the upstream stator cannot always remain away from the rotor blade because of its rotation. As a result, the temperature distributions over the rotor blade surfaces are significantly affected. 
In case 4N, two streamwise low-temperature bands appear on each rotor blade surface, corresponding to the two layers of cold fuel jets introduced at the turbine inlet.
In contrast, the two cold bands become high-temperature regions in case 4R as the injected fuel is burned in the turbine passage. Moreover, the bands on the pressure surface are wider and stronger than those on the suction surface. This is because the hot streaks originating from the upstream flames in the stator directly impinge on the rotor pressure surface, as shown in Fig. \ref{fig:contour_instant_radial_slice}. Consequently, the pressure-surface temperature in case 4R remains high, exceeding $2050 \, \mathrm{K}$ in some local locations.
In case 16R, the hot bands on the rotor surfaces spread out more evenly across the blade height, significantly lowering the danger of hot streaks found in case 4R, with the maximum value not exceeding $1900 \, \mathrm{K}$. This is consistent with the flatter total temperature profiles at both the stator and rotor outlets in case 16R, as shown in Figs. \ref{fig:spandistr_T0_stator_outlet} and \ref{fig:spandistr_T0_rotor_outlet}.

High temperature levels over the rotor blade surfaces in the reacting cases cause a challenge to the blade cooling. This preliminary study demonstrates that combustion in a turbine stage can be managed to allow additional fuel burn without causing excessive temperature rise on either the stator and rotor blades.
Notice that adiabatic boundary conditions are applied on all blade surfaces, and the design of the fuel-injector system has not been optimized in this study. 
Further improvement in heat management can be achieved by optimizing and controlling the fuel-injector system in addition to applying conventional blade cooling. The temperature levels over turbine blades can also be relieved by reducing the fuel supply to the primary combustor, thereby lowering the turbine inlet temperature, since the turbine-burner is capable of undertaking part of workload of the primary combustor. These issues remain as topics for future investigation.

\subsection{Thermodynamic Performance} \label{sec:thermodynamic_performance}
\begin{figure}[htb!]
    \centering
    \includegraphics[width=0.4\linewidth]{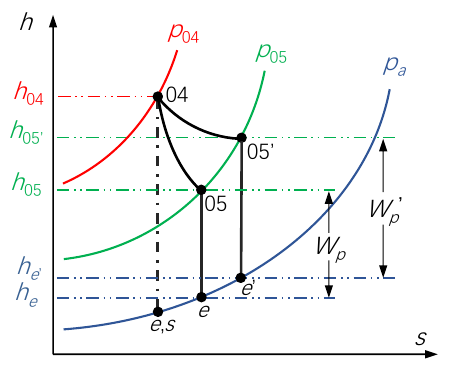}
    \caption{Enthalpy-entropy diagram of a turbine section.}
    \label{fig:h-s_thermodyanmic_process}
\end{figure}

This section assesses the benefits of a turbine-burner by examining the thermodynamic process of the turbine section in a gas-turbine engine. The turbine section generally consists of an HPT, an LPT, followed by a possible propulsion nozzle. 
Consider the enthalpy-entropy ($h$-$s$) diagram of a turbine section in which the present turbine stage serves as the first stage of the HPT, as shown in Fig. \ref{fig:h-s_thermodyanmic_process}. 
Given the total enthalpy $h_{04}$ and pressure $p_{04}$ at the turbine inlet and the ambient pressure $p_a$ for the engine, the ideal amount of work per unit mass (including the kinetic energy of the exhaust gas) that can be extracted is determined by the enthalpy difference $h_{04} - h_{e,s}$, where $h_{e,s}$ is the enthalpy of the flow when expanded to the ambient pressure isentropically, as shown by the vertical dot-dashed line in Fig. \ref{fig:h-s_thermodyanmic_process}. 

If we insert our conventional E\textsuperscript{3} high-pressure turbine stage to the ($h_{04}, p_{04}$) state and have it expand to $p_{05}$, the process will be non-isentropic, resulting in an outlet total enthalpy $h_{05}$. The turbine work extracted from the flow can then be computed by $W_t = h_{04} - h_{05} = P_t/\dot{m}$.
The gas at the exit of this turbine stage is still at high pressure and temperature, and therefore can be used to drive another turbine (LPT) for additional work and/or expand through a nozzle to ambient pressure to obtain kinetic energy of the gas for thrust. We characterize this capacity of the gas by the ideal residual work $W_p$ available by expanding the gas isentropically to the ambient pressure along the vertical line $05$-$e$ in Fig. \ref{fig:h-s_thermodyanmic_process}.

In the turbine-burner case, more work is extracted out of the turbine with the same pressure drop to $p_{05}$ because of the heat addition. Let $W_t^\prime$ be the new turbine work. We define the rate of work increase due to the use of the turbine-burner as
\begin{equation} \label{eq:work_rt}
    r_t = \frac{\Delta W_t}{W_t} = \frac{W_t^\prime - W_t}{W_t} = \frac{W_t^\prime}{W_t} - 1
\end{equation}
The total enthalpy at the turbine-stage outlet will be higher, as marked by $h_{05^\prime}$ in Fig. \ref{fig:h-s_thermodyanmic_process}, which will result in a higher residual work $W_p^\prime$. 
By assuming a constant specific heat capacity $C_p = 1300 \, \mathrm{J/kg/K}$ and specific heat ratio $\gamma = 1.29$, the residual works $W_p$ and $W_p^\prime$ can be computed by
\begin{subequations}
    \begin{align}
        W_p &= h_{05}- h_{e} = C_pT_{05}\left(1-\pi_p^{-\frac{\gamma-1}{\gamma}}\right) \\ 
        W_p^\prime &= h_{05}^\prime- h_{e}^\prime = C_pT_{05}^\prime \left(1-\pi_p^{-\frac{\gamma-1}{\gamma}}\right)
    \end{align}
\end{subequations}
where $\pi_p = p_{05}/p_a = 12.8$ is the pressure ratio within the processes $05$-$e$ and $05^\prime$-$e^\prime$. 
The residual work increase rate due to combustion is thus defined as
\begin{equation} \label{eq:work_rp}
    r_p = \frac{\Delta W_p}{W_p} = \frac{W_p^\prime - W_p}{W_p} = \frac{T_{05}^\prime}{T_{05}} - 1
\end{equation}
Considering the thermodynamic process within the entire turbine section, we introduce the overall work increase rate
\begin{equation} \label{eq:work_ro}
    r_o = \frac{\Delta W_t + \Delta W_p}{W_t + W_p} = \frac{W_t^\prime + W_p^\prime}{W_t + W_p} - 1 
\end{equation}
By this definition, the value of $r_o$ should be between those of $r_t$ and $r_p$.

The additional turbine work and residual work are obtained at the expense of the additional fuel. Therefore, we define a thermal efficiency for the turbine stage
\begin{equation}
    \eta_f = \frac{\Delta W_t + \Delta W_p}{E_f} 
\end{equation}
where $E_f = Y_f \cdot \mathrm{LHV}$ is the thermal energy of the fuel added to the turbine stage per unit mass flow. Here, $Y_f$ is the overall mass fraction of CH\textsubscript{4} at the turbine inlet, and LHV is the lower heating value of CH\textsubscript{4}, taken as $50 \, \mathrm{MJ/kg}$.

Table \ref{tab:overall_performance_parameter} summarizes the thermodynamic performance of the turbine stage. The three work increase rates, $r_t$, $r_p$, and $r_o$, as well as $\eta_f$, for the two reacting cases 4R and 16R are defined relative to case 0N. 
The outlet temperatures in the two reacting cases are higher than those in the nonreacting cases, with case 16R yielding a slightly lower value than case 4R. These trends are consistent with the overall enthalpy fluxes in Table \ref{tab:overall_energy_balance_turbine} and the relative total temperature in Fig. \ref{fig:T0-x_nonreacting_reacting}.
The outlet pressures are similar in both the nonreacting and reacting cases, with relative differences below 1.3\%. This supports the assumption of using the same outlet pressure $p_{05}$ in both nonreacting and reacting processes in Fig. \ref{fig:h-s_thermodyanmic_process}.

\begin{table}[htb!]
    \centering
    \caption{Thermodynamic performance of the turbine stage}
    \begin{tabular}{cccccccc}
        \hline
        Case & $T_{05}$, K & $p_{05}$, bar & $W_t$, kJ/kg & $r_t$, \% & $r_p$, \% & $r_o$, \% & $\eta_f$, \% \\  
        \hline
        0N  & 1379 & 12.71 & 359.1 & -- & -- & -- & --          \\ 
        4N  & 1360 & 12.69 & 353.7 & -- & -- & -- & --          \\ 
        4R  & 1617 & 12.86 & 389.5 &  8.5 & 17.3 & 14.5 & 44.1  \\ 
        16R & 1599 & 12.82 & 400.4 & 11.5 & 16.0 & 14.6 & 44.3  \\ 
        \hline
    \end{tabular}
    \label{tab:overall_performance_parameter}
\end{table}

The turbine work per unit mass is increased by 8.5\% in case 4R and 11.5\% in case 16R, as indicated by the values of $r_t$.  
Although we may not need the extra work output by the turbine for a given compressor pressure and the core engine mass flow, we can scale-down the turbine size or number of stages to provide the same power to the compressor. A better design change is to perform a re-design of the whole engine cycle to take advantage of both the increased turbine work and the increased available residual work to achieve an engine with much higher thrust for the same size or an engine with reduced size and weight for the same thrust. The latter also has the benefits of reduced installation drag and weight. For turbofan engines, there is even more potential for the turbine-burner for high thrust and high efficiency at the same time as the extra power of the turbine-burner and available residual power may be used to drive an ultra-high bypass fan and the core compressor stages with higher overall pressure ratios. Such topics were briefly explored in Refs. \cite{sirignano1999performance, liu2001turbojet}. As we advance our understanding and technology of the turbine-burner, system-level design and optimization can be performed for missions that have not been possible before.

According to Eq. (\ref{eq:work_rp}), the residual work increases with increasing total temperature $T_{05^\prime}$ at the turbine-stage outlet for a given total pressure $p_{05^\prime}$. The premise of the turbine-burner engine \cite{liu2001turbojet, sirignano1999performance} is to balance the fuel addition with work extraction so that the total temperature is kept constant across a turbine stage within the material limit of the turbine blades while maximizing power output. 
Hence, a value of $T_{05^\prime}$ equal to or slightly less than the turbine inlet total temperature $T_{04}$ is desired. In the present study with the E\textsuperscript{3} HPT turbine design, we obtain
$1617 \, \mathrm{K}$ and $1599 \, \mathrm{K}$, yielding residual work increase rates of 17.3\% and 16.0\% in cases 4R and 16R, respectively. Both of them are higher than their turbine-stage work increase rates. 
Although the values of $r_t$ and $r_p$ are different in cases 4R and 16R, both cases exhibit nearly identical overall work increase rate $r_o$ and thermal efficiency of fuel burning $\eta_f$. The overall work increases by 14.5\%. The thermal efficiency $\eta_f$ is approximately 44\%, which is higher than the overall thermal efficiency of modern gas-turbine engines, which ranges from 30\% to 43\%.
Note that this thermal efficiency for the additional fuel burn is calculated without accounting for the component efficiencies of the downstream turbine or nozzle.

We analyze the work extraction process of a turbine-burner from both thermodynamic and mechanical views to provide guidance for design. Given the turbine inlet total temperature $T_{04}$ and the total pressure ratio $\pi_t = p_{04}/p_{05}$ (see Fig. \ref{fig:h-s_thermodyanmic_process}), the work from an ideal conventional process is 
\begin{equation} \label{eq:isentropic_work}
    W_t = -\int_{04}^{05} \frac{1}{\rho} \mathrm{d}p = \frac{\gamma}{\gamma-1}RT_{04}\left(1-\pi_t^{-\frac{\gamma-1}{\gamma}}\right) 
\end{equation}
where the isentropic relation and perfect-gas law are assumed. For the ideal turbine-burner, heat is added while an equivalent amount of work is extracted to maintain an isothermal process. This requires $h_{05^\prime} = h_{04}$ and 
\begin{equation} \label{eq:isothermal_work}
    W_{t}^\prime = -\int_{04}^{05^\prime} \frac{1}{\rho} \mathrm{d}p = RT_{04} \ln{\pi_t} = \frac{\dot Q_t}{\dot m}
\end{equation}

\begin{figure}[htb!]
    \centering
    \includegraphics[width=0.49\linewidth]{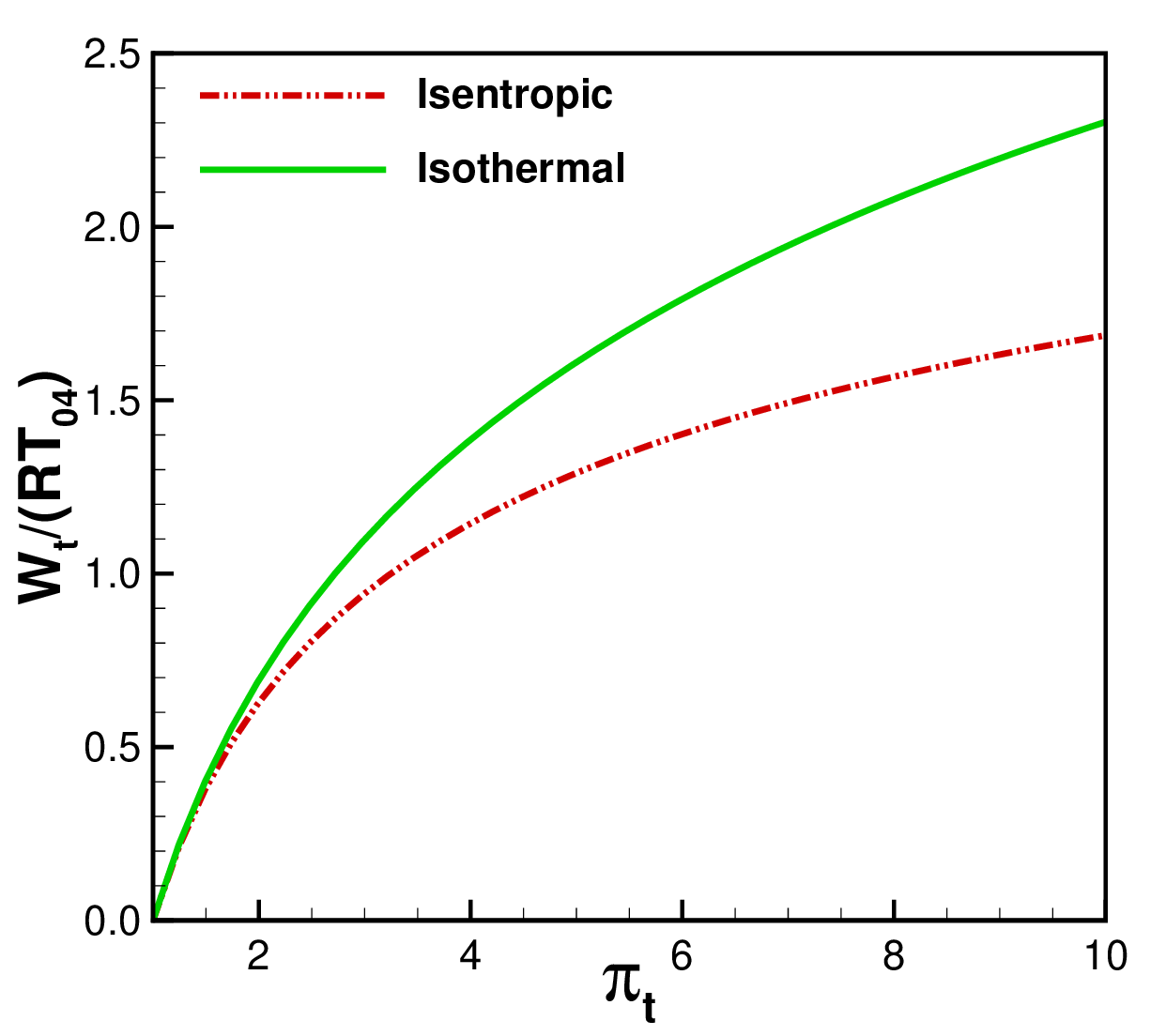}
    \caption{Variation of theoretical turbine work with pressure ratio.}
    \label{fig:theoretical_work_isentropic_isothermal_process}
\end{figure}

Figure \ref{fig:theoretical_work_isentropic_isothermal_process} shows the variation of the theoretical turbine work given by Eqs. (\ref{eq:isentropic_work}) and (\ref{eq:isothermal_work}) with the pressure ratio. The theoretical work in the isothermal process is higher than that in the isentropic process, and the difference between them rises with the turbine pressure ratio. Specifically, at the pressure ratio of 2.34 in the present turbine stage, the relative difference is 10\%. Recall from Table \ref{tab:overall_performance_parameter} that the turbine-stage work increase rates in cases 4R and 16R are 8.5\% and 11.5\%, respectively. Both of them are close to the theoretical value. 
Equations (\ref{eq:isentropic_work}) and (\ref{eq:isothermal_work}) also explain why the turbine work was nearly unchanged between the reacting and nonreacting cases of an experimental low-speed turbine stage in the previous study \cite{zhu2025_turbine_burner}. Its total pressure ratio is 1.25, which gives a theoretical work increase rate of only 2.5\%.
To raise the gain in the work output in a turbine-burner, a higher pressure ratio is desirable as pointed out in Refs. \cite{sirignano1999performance, liu2001turbojet}.

From a mechanical point of view, the turbine work per unit mass within one rotor passage is given by the Euler work equation for turbomachinery
\begin{equation} \label{eq:euler_work_equation}
    W_t = U(V_{\theta 4.5} - V_{\theta 5}) = U(V_{\theta 4.5} - V_{\theta r 5} - U) = U^2 \left[ \frac{V_{x4.5}}{U} \left( \tan \beta_{4.5} - \frac{V_{x5}}{V_{x4.5}} \tan \beta_{r5} \right) - 1 \right]  
\end{equation}
where $U$ is the rotor blade velocity; $V_\theta$ and $V_x$ are the circumferential and axial components of the face-averaged flow velocity, respectively, with $V_\theta = V_x \tan \beta$. The subscripts, $4.5$ and $5$, denote the stator and rotor outlets, respectively. The subscript $r$ represents the quantities measured in the rotor-relative reference frame. The flow angles are positive in the rotor direction.

Equation (\ref{eq:euler_work_equation}) implies that the turbine work can be improved by raising $U$, $V_{x}/U$, and $\beta_{4.5}$, or by reducing $\beta_{r5}$ ($\beta_{r5} < \beta_{4.5}$ and is negative). 
$U$ is directly determined by the turbine geometry and the rotation speed $\mathbf{\Omega}$. 
$\beta_{4.5}$ is the turning angle of the flow by the stator blade and $\beta_{r5}$ is the relative flow angle at the rotor outlet. Once the blade geometry is designed, the flow is expected to follow the blade metal angles with little deviation for well-behaved flow. Adding heat to the blade passage has little impact on these flow angles, as shown by the results in Figs. \ref{fig:spandistr_beta_stator_outlet} and \ref{fig:spandistr_beta_rotor_outlet}.
Although not shown here, the axial velocities $V_{x4.5}$ and $V_{x5}$ are higher in the reacting cases, with those in case 16R being the highest. Both produce a higher value of $W_t$, explaining why combustion enhances the turbine-stage work, as reported in Table \ref{tab:overall_performance_parameter}.
To fully enhance the work extraction potential of a turbine-burner, it is necessary to  redesign the turbine geometry for an improved velocity triangle within the turbine stage.

The above analysis demonstrates the necessity of re-designing the blade rows for a turbine-burner stage with a coupled consideration of aerodynamic (mechanical) and thermodynamic aspects. An aerodynamically under-designed blade configuration may yield a turbine stage not able to extract the amount of work needed to keep the temperature constraint, yielding excess heat that may damage the blades. On the other hand, an over-designed blade configuration may yield lower exit temperature that does not take full advantage of the turbine-burner concept. In addition, optimization of the pressure ratios, maximum temperature, and multi-staging must be considered in the context of engine performance.

\section{Conclusions} \label{sec:conclusion}
The CTB design for a gas-turbine engine has the potential for significant increase in specific thrust/power with higher thermal efficiency. To put it into practice, however, requires investigation into the possibility of stable and efficient combustion within a rotating turbine stage without excessive heating to the blades and loss of aerodynamic efficiency.
The present study applied LES to compute chemically reacting flow in a practical turbine stage and to analyze the influence of fuel injection and combustion on its aerodynamic and thermodynamic performance. Four cases were considered: case 0N with uniform vitiated-air inlet, cases 4N and 4R with four fuel injectors at the inlet, and case 16R with sixteen smaller fuel injectors. The first two cases are nonreacting, while the latter two involve combustion. The turbine-stage analyses indicate viability for the turbine-burner concept. 

The reacting flow field in case 4R was examined.
In the stator, tubular-like flames are established between fuel and air and convected downstream near its suction and pressure surfaces. The suction-surface flame becomes weaker in the stator passage until reaching the trailing edge due to the strong favorable streamwise pressure gradient, while the pressure-surface flame quenches at the stator mid-chord as the fuel has been depleted by the strong reactions in the upstream. The suction-surface flame is enhanced again in the stator wake and then quenches slightly downstream of the rotor leading edge due to fuel depletion. 
The flames and hot streaks from the stator impinge on the rotor pressure surface and then are convected towards the suction surface of the adjacent rotor blade. 

The influence of fuel injection and combustion on the aerodynamic performance of the turbine stage was assessed. 
The small amount of fuel injected at the inlet has minimal influence on the total-pressure loss within the turbine stage.
The turbine powers in the reacting cases are higher than those in the nonreacting cases, attributed to the effects of combustion on the static pressure over the rotor blade surfaces.
However, the mass flow rates in cases 4R and 16R decrease by 7\% and 8\% relative to the nonreacting cases, respectively.
Compared with case 4R, case 16R exhibits earlier and faster chemical reactions, as well as more uniform spanwise distributions of total temperature at the stator and rotor outlets.

Four performance parameters were introduced to assess the work increase potential of the turbine-burner. 
Compared with the baseline case 0N, the turbine-stage work increases by 8.5\% in case 4R and by 11.5\% in case 16R, while the residual work rises by 17.3\% and 16.0\%, respectively. 
The two reacting cases exhibit nearly identical overall work increase rates of 14.5\% and achieve a thermal efficiency of 44\%, which is generally comparable to the overall thermal efficiency of modern gas-turbine engines. 

The mechanisms of turbine work improvements through combustion were revealed by examining the ideal thermodynamic processes and the Euler work equation within the turbine stage.
These analyses from both thermodynamic and mechanical views provide guidance for turbine-burner design with improved performance.
(1) Early completion of combustion in the upstream stator can enhance the work output in the downstream rotor. 
(2) Raising the turbine pressure ratio can improve the work extraction of a turbine-burner. 
(3) Combustion alters the axial velocity but not the stator-outlet flow angle. Achieving the full work-extraction potential of a turbine-burner demands a redesigned or modified turbine geometry.

Combustion in a turbine stage can be managed to allow additional fuel burn without causing excessive temperature rise on either the stator and rotor blades. 
Fuel injection and chemical reaction have little effects on the temperature over the stator blade surface since the spray from the fuel injectors can be arranged to pass through the stator passage away from the blade surfaces. Local high temperature on the rotor blade can be suppressed by using a more uniform spanwise distribution of fuel injectors, as verified by case 16R. 
In future, further improvement in heat management can be achieved by optimizing and controlling the fuel-injector system in addition to applying conventional blade cooling.

\section*{Acknowledgments}
This research was supported by the Office of Naval Research through Grant N00014-22-1-2467 with Dr. Steven Martens as program manager.
This research was granted access to the high-performance computing resources of the U.S. Department of Defense. 
The authors acknowledge Sylvain L. Walsh of University of California Irvine for valuable discussions.

\bibliography{main}

\end{document}